\documentclass[conference]{IEEEtran}

\usepackage{booktabs}
\usepackage{paralist}
\usepackage[group-separator={,}, per-mode=symbol, binary-units]{siunitx}
\usepackage{xspace}
\usepackage{eurosym}
\usepackage[dvipsnames]{xcolor}

\usepackage{graphicx}
\usepackage{color, colortbl}
\usepackage[ruled,vlined]{algorithm2e}
\SetKwComment{Comment}{$\triangleright$\ }{}
\usepackage{mathtools}
\DeclarePairedDelimiter\ceil{\lceil}{\rceil}

\usepackage{amsfonts}
\usepackage{amsmath}
\usepackage{bm}
\usepackage{threeparttable, tablefootnote}
\usepackage{multirow}
\usepackage{lipsum}                     %
\usepackage{soul}
\usepackage{comment}

\newcommand{\ie}{i.\@\,e.\@\xspace}
\newcommand{\eg}{e.\@\,g.\@\xspace}
\newcommand{\cf}{cf.\@\xspace}
\newcommand{\etal}{et~al.\@\xspace}

\newcommand{\Paragraph}[1]{\smallskip\noindent{\bf #1.}}

\definecolor{Gray}{gray}{0.5}
\PassOptionsToPackage{hyphens}{url}\usepackage{hyperref}

\ifCLASSOPTIONcompsoc
  \usepackage[caption=false,font=normalsize,labelfont=sf,textfont=sf]{subfig}
\else
  \usepackage[caption=false,font=footnotesize]{subfig}
\fi

\begin{document}

\title{IRShield: A Countermeasure Against Adversarial Physical-Layer Wireless Sensing}

\author{\IEEEauthorblockN{
Paul Staat\IEEEauthorrefmark{1},
Simon Mulzer\IEEEauthorrefmark{2}, 
Stefan Roth\IEEEauthorrefmark{2},
Veelasha Moonsamy\IEEEauthorrefmark{2},\\
Markus Heinrichs\IEEEauthorrefmark{3},
Rainer Kronberger\IEEEauthorrefmark{3},
Aydin Sezgin\IEEEauthorrefmark{2},
Christof Paar\IEEEauthorrefmark{1}
}
\IEEEauthorblockA{\IEEEauthorrefmark{1}Max Planck Institute for Security and Privacy, Bochum, Germany}
\IEEEauthorblockA{\IEEEauthorrefmark{2}Ruhr University Bochum, Bochum, Germany}
\IEEEauthorblockA{\IEEEauthorrefmark{3}TH Köln – University of Applied Sciences, Cologne, Germany}
\IEEEauthorblockA{E-mail: \{paul.staat, christof.paar\}@mpi-sp.org, simon@mulzer.eu, \{stefan.roth-k21, aydin.sezgin\}@rub.de, \\email@veelasha.org, \{markus.heinrichs, rainer.kronberger\}@th-koeln.de}
}

\maketitle
\thispagestyle{plain}
\pagestyle{plain}

\begin{abstract}

Wireless radio channels are known to contain information about the surrounding propagation environment, which can be extracted using established wireless sensing methods. Thus, today's ubiquitous wireless devices are attractive targets for passive eavesdroppers to launch reconnaissance attacks. In particular, by overhearing standard communication signals, eavesdroppers obtain estimations of wireless channels which can give away sensitive information about indoor environments. For instance, by applying simple statistical methods, adversaries can infer human motion from wireless channel observations, allowing to remotely monitor premises of victims. In this work, building on the advent of intelligent reflecting surfaces (IRSs), we propose IRShield as a novel countermeasure against adversarial wireless sensing. IRShield is designed as a plug-and-play privacy-preserving extension to existing wireless networks. %
At the core of IRShield, we design an IRS configuration algorithm to obfuscate wireless channels. We validate the effectiveness with extensive experimental evaluations. In a state-of-the-art human motion detection attack using off-the-shelf \mbox{Wi-Fi} devices, IRShield lowered detection rates to~\SI{5}{\percent} or less.

\end{abstract}

\IEEEpeerreviewmaketitle

\section{Introduction}

Wireless connectivity drives many current digital innovations %
and is becoming increasingly ubiquitous. This trend manifests itself in a worldwide surge in the adoption of Internet of Things (IoT) devices with $75$~billion connected  devices projected by 2025~\cite{iotadoption}. The IoT phenomenon is already heavily present in our daily lives: voice assistants, watches, locks, light bulbs, cameras, vacuum cleaners, sensors, and actuators have turned into devices that we consider \textit{smart}. However, virtually all of them rely on wireless connectivity, based on standards such as \mbox{Wi-Fi}, BLE, NB-IoT, or ZigBee~\cite{wirelessconnectivity2025}. 

A downside of having ubiquitous wireless communication is the burgeoning of new types of privacy concerns. %
Since wireless communication is based on an open medium, it is inherently shared with third, potentially adversarial, parties. Although cryptographic primitives are widely used for providing confidentiality and integrity of the  application data, passive eavesdroppers are still able to exploit sensitive information from sniffed radio frequency~(RF) signals. This is possible as the propagation of RF signals depends on the physical surroundings of devices, \eg, due to reflections off walls, objects, and individuals. Eavesdroppers can remotely observe such propagation effects to gather insights about the physical environment of a legitimate transmitter, referred to as (adversarial) wireless sensing. %
As already demonstrated by existing work in the literature, this leads to an immediate privacy threat due to reconnaissance attacks, such as adversarial human motion sensing. %

Zhu~\etal~\cite{zhu.2020} demonstrated the feasibility of carrying out fine-grained detection and tracking of human movements inside a building by only sniffing ordinary \mbox{Wi-Fi} signals. Their low-cost attack makes use of the \mbox{Wi-Fi} signal dynamics and variance of multipath signal propagation to track indoor movements without requiring any a priori knowledge on the adversary's side. Similarly, Banerjee~\etal~\cite{banerjeeViolating.2014} proposed a methodology to detect human motion %
within the line-of-sight~(LOS), \ie, the direct path between the victim's device and that of the passive sniffer. Generally, there is significant potential for (adversarial) wireless sensing on the physical layer, as highlighted by the large corpus of existing work surveyed by Ma~\etal~\cite{ma.2019}. Besides exploiting physical-layer properties, an adversary can also monitor packet-level information to infer the state of IoT devices and actions carried out by users present in the network, cf.~\cite{peekaboo, Trimananda.2020, Zhang.2018}. 

While inference attacks from packet-level wireless sniffing may be thwarted by means of phantom users~\cite{liu.2021}, a versatile and easy to use countermeasure against wireless sensing on the physical layer is still lacking. One prominent approach was put forward by Qiao~\etal who proposed PhyCloak~\cite{qiao.2016}. By using a full-duplex radio, ambient wireless signals are re-transmitted with randomized modifications to obfuscate sensitive physical information. %
However, full-duplex is costly as it requires specialized and complex radio hardware. While eliminating the use of full-duplex equipment, the proposal by Yao~\etal~\cite{yao.2018} requires several motorized moving hardware parts. Jiao~\etal~\cite{jiao.2021} proposed modifying software-defined wireless transmitters. Other approaches by Zhu~\etal~\cite{zhu.2020} and Wijewardena~\etal~\cite{wijewardena.2020} affect the allocation of the channel. Therefore, they not only trade the quality of service of wireless communication against adversarial sensing capabilities, but also require integration with the wireless devices in the field.

In this paper, we aim to design a countermeasure against adversarial wireless sensing on the physical layer that overcomes the shortcomings of previous approaches and resolves the following challenges (C1--C3):%

\begin{compactitem} 
    \item \textbf{Device-agnostic countermeasure (C1)}: There exists a wide variety of wireless devices that are present in the real-world and cannot be remotely updated or modified. Thus, we pursue a solution  independent of deployed devices, the used wireless waveforms, and standards.  %
    \item \textbf{Maintain quality of service (C2)}: The connectivity requirements of wireless applications are diverse and hard to predict, which renders a reduction in quality of service unacceptable. An effective countermeasure should therefore not affect the quality of the wireless link.
\end{compactitem}

Inspired by a recent trend from the realm of wireless communication, we explore \textit{smart radio environments} with intelligent reflecting surfaces (IRSs) to overcome these hurdles. To date,  digitally controlled IRS are primarily used as adjustable elements in propagation environments, which improve wireless communication~\cite{Renzo2019}. A key aspect of the IRS is that it directly affects the adversary's observation used for privacy-violation attacks: the wireless channel. Therefore, we propose \textit{IRShield} -- a practical IRS-based wireless channel \textit{obfuscation}. It greatly reduces adversarial capabilities for privacy violations from  wireless sensing. Specific to the IRS, we address the  third challenge:

\begin{compactitem}
    \item \textbf{Surface configuration (C3)}: In a communication context, IRS are configured based on channel information from legitimate receivers and careful integration into the wireless infrastructure~\cite{wuSmartReconfigurableEnvironment2020}. However, in an adversarial wireless sensing scenario, the receiver (eavesdropper) is unknown and hostile, and therefore the required channel information is not available. The challenge is thus to provide strong channel obfuscation, given the extremely large IRS configuration space of, \eg, $2^{256}$.
\end{compactitem}

IRShield addresses \textbf{C3} through a probabilistic surface configuration strategy specifically designed to achieve channel obfuscation as a standalone application. Further, IRShield affects the bare radio wave propagation, regardless of specific devices or waveforms, thus meeting the criteria for addressing \textbf{C1}. Finally, IRShield does not allocate the wireless channel itself but affects it randomly, hence solving \textbf{C2}.

\Paragraph{Contribution} %
We are the first to propose IRS to be used as a hands-on countermeasure against unauthorized wireless sensing. We design a dedicated algorithm to generate randomized IRS configurations to achieve wireless channel obfuscation. Our scheme, which we refer to as IRShield, is laid out as a standalone privacy-preserving extension for plug-and-play integration with existing wireless infrastructure. %
Finally, we present an extensive experimental evaluation of the proposed technique, showing that it successfully prevents adversarial motion sensing. In particular, IRShield lowered detection rates to~\SI{5}{\percent} or less in a state-of-the-art attack on \mbox{Wi-Fi} devices. Our measurement data is available online in order to support research reproducibility~\cite{meas_data_reference}.

\section{Background}\label{sec:background}

We provide technical background on wireless sensing methods and smart radio environments, including IRS. For the reader's convenience, we have summarized the important terminology and symbols in Table~\ref{tab:terminology}.

\label{sec:terminology_overview}
\begin{table}[t]
\footnotesize
\centering
\caption{Terminology overview}
\label{tab:terminology}
    \begin{tabular}{|c|l|}
        \hline $t$ & Time index\\
        \hline $\mathcal{L}_t$ & Propagation paths at time $t$\\
        \hline $K$ & Number of subcarriers \\
        \hline $k$ & Subcarrier index\\
        \hline $N_{\mathrm{T}}$ & Number of transmit antennas \\
        \hline $N_{\mathrm{R}}$ & Number of receive antennas\\
        \hline $M$ & Number of IRS elements \\
        \hline $r_m(t)$ & IRS reflection coefficient\\
        \hline $\bm{H}(k,t)$ & Channel at subcarrier $k$ \\
        \hline $\bar{\sigma}^{(w)}(t)$ & Average channel standard deviation\\
        \hline
    \end{tabular}
\end{table}

\subsection{Wireless Sensing Methods}\label{sec:background-wsensing}
\begin{figure}
\centering
\includegraphics[width=0.7\linewidth]{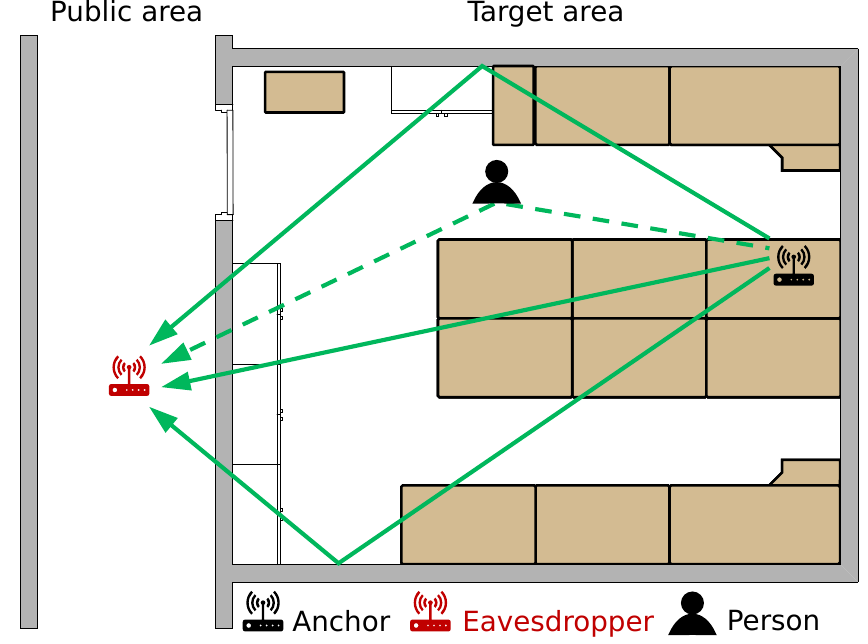}
\caption{Illustration of wireless signal propagation in indoor environments.} %
\label{fig:floorplan_motion_sensing_propagation}
\end{figure}

When wireless radio signals traverse from a transmitter to a receiver, they are distorted by the \textit{channel response} which aggregates environment-dependent effects on the signal such as multipath propagation. %
In particular, the channel response between an arbitrary transmitter and receiver %
is determined by different propagation paths through the environment, as illustrated in \figurename~\ref{fig:floorplan_motion_sensing_propagation}. These are LOS paths and various non-LOS (NLOS) paths, \eg, from reflections off walls, interior objects, or individuals. Particular to the case of human motion, new paths are created and existing paths are blocked, thus making the wireless signal propagation time-variant. At time $t$, we assume that the propagation paths $\mathcal{L}_t$ are present. 

In a wireless communication context, the (undesired) channel response is regularly estimated from known preamble sequences to be compensated subsequently from the received signal. %
This, likewise, happens for every received packet when considering IEEE~802.11n \mbox{Wi-Fi} communication based on orthogonal frequency division multiplexing (OFDM). OFDM divides a wide bandwidth wireless channel into $K$~parallel independent (\ie, orthogonal) narrowband channels, \ie, subcarriers, for data transmission. Upon detection of a packet, a \mbox{Wi-Fi} receiver estimates the channel response (also referred to as channel state information~(CSI)) for each subcarrier $k$ using a known preamble to obtain
\begin{align}
    \bm{H}(k,t)=\underbrace{\sum_{l\in\mathcal{L}_t}^{}\bm{G}_l(k,t)}_{\substack{\text{propagation paths}\\\text{through surrounding}}}+\underbrace{\vphantom{\sum_{l\in\mathcal{L}_t}^{}}\bm{N}(k,t)}_{\text{noise}}.
    \label{eq:channel_multipath}
\end{align}
Considering a multiple-input multiple-output (MIMO) system typical for \mbox{Wi-Fi} in which the transmitter or receiver has $N_\mathrm{T}$ or $N_\mathrm{R}$ antennas, $\bm{G}_l(k,t)$ represents the $N_{\mathrm{R}}\times N_{\mathrm{T}}$-dimensional signal propagation matrix of the $l^{th}$ propagation path. This matrix then involves the complex channel gain and possible steering or response vectors at a certain time $t$. $\bm{N}(k,t)$ denotes measurement noise.

Building upon such channel estimates from conventional communication systems, \eg, \mbox{Wi-Fi}~\cite{ma.2019}, \textit{wireless sensing} seeks to extract information on the physical propagation environment contained in the channel response. %
Applications such as human activity and gesture recognition, imaging, or vital sign monitoring~\cite{yousefiSurveyBehaviorRecognition2017, ma.2019} impressively demonstrate the ability of commodity wireless devices to detect environmental conditions and provide novel use-cases for the communication equipment. However, despite its many legitimate uses, wireless sensing can also be deployed by adversaries: Passive eavesdroppers can receive packets and gain information about remote environments, leading to potential privacy violations of victim parties.

When it comes to the threat of adversarial human motion sensing, it is, first and foremost, important to understand the mechanisms underlying the attack. First, the amplitude of $\bm{G}_l(k,t)$ depends on whether a path is temporarily blocked by a person at time $t$ or not. Further, motion can introduce temporary new propagation paths as the human body also reflects signals. Thus, the impact of human motion on the wireless channel is expected to vary over time, depending on the individual's position and the number of affected propagation paths. Further, due to distance-dependent path loss of radio wave propagation, shorter propagation paths, \eg, LOS or short NLOS paths, contribute most to the channel. This is an important aspect, since a person moving into the LOS affects (blocks) the strongest propagation paths, yielding strong variation of the wireless channel.

\subsection{Smart Radio Environments}
\label{sec:background_irs}

The IRS is considered a promising technology for future wireless networks~\cite{Liu:2021wy} as it enables \textit{smart radio environments}~\cite{Renzo2019}. More precisely, the IRS is a synthetic surface that has digitally reconfigurable reflection properties of radio waves. This rather new concept is rooted in physics research on metamaterials and metasurfaces~\cite{kaina_shaping_2015}, which recently saw drastic simplification that led to the IRS. Here, many reflecting elements distributed across a surface are individually and electronically adjustable, allowing dynamic manipulation of impinging radio waves. For instance, the IRS may be configured to reflect signals into a particular direction. Thus, the IRS allows partial reconfiguration of the radio propagation environment to alter the wireless channel.

We consider an IRS consisting of $M$ reflecting elements with the $m^{\textrm{th}}$ configurable reflection coefficient $r_m(t)$. The reflection coefficient can be selected among a discrete set of values and is changing the reflected signal, such that the incoming wave is multiplied by this factor. For each element, additional propagation paths become part of the channel. Incorporating the IRS contribution, we rewrite Eq.~\eqref{eq:channel_multipath} as
\begin{align}
    \bm{H}(k,t)=\underbrace{\sum_{l\in\mathcal{L}_t}^{}\bm{G}_l(k,t)}_{\substack{\text{propagation paths}\\\text{through surrounding}}}+\underbrace{\vphantom{\sum_{l\in\mathcal{L}_t}^{}}\sum_{m=1}^{M}r_m(t)\bm{G'}_m(k,t)}_{\substack{\text{propagation paths}\\\text{via IRS}}}+\underbrace{\vphantom{\sum_{l\in\mathcal{L}_t}^{}}\bm{N}(k,t)}_{\text{noise}},
    \label{eq:channel_multipath_irs}
\end{align}
where $\bm{G'}_m(k,t)$ is the signal propagation via the $m^{\textrm{th}}$~IRS element. %
Eq.~\eqref{eq:channel_multipath_irs} shows how the channel matrices can be modified by engineering the values of $r_m(t)$. On the other hand, randomizing the IRS configuration implies partial randomization of the channel.%

\section{Attack Scenario}
\label{sec:attack}

In this section, we introduce the threat model and outline the adversary's strategy, as described in the state-of-the-art attack in~\cite{zhu.2020}. Further, we elaborate on the experimental evaluation. %

\subsection{Threat Model}
\label{sec:threat_model}

In this work, we consider a number of legitimate \mbox{Wi-Fi} devices which transmit packets. The devices, which we also refer to as \textit{anchors}, are deployed within an ordinary environment, such as a home or an office. The goal of the passive eavesdropper, Eve, is to infer human motion within the environment by analyzing eavesdropped packets. We assume that Eve possesses a wireless receiver that is able to pick up and demodulate signals originating from the anchors. Eve does not have access to the environment but can position her receiver at arbitrary public locations outside the perimeter. After initial positioning of the receiver, Eve can then proceed to act remotely. Furthermore, we assume Eve is not able to break the applied cryptography, \ie, to read secured payload data. However, since the anchors employ standard-compliant \mbox{Wi-Fi} communication, Eve can obtain physical-layer channel estimations from the known packet preambles. 

We assume that the owner (i.e. the \textit{defender}) of the environment can position the anchor devices at will. Furthermore, we assume the defender can place one or multiple IRS within their space and apply customized configurations.

\subsection{Eavesdropper Strategy as per~\cite{zhu.2020}}
\begin{figure}
\centering
\subfloat[]{\includegraphics[width=0.49\linewidth]{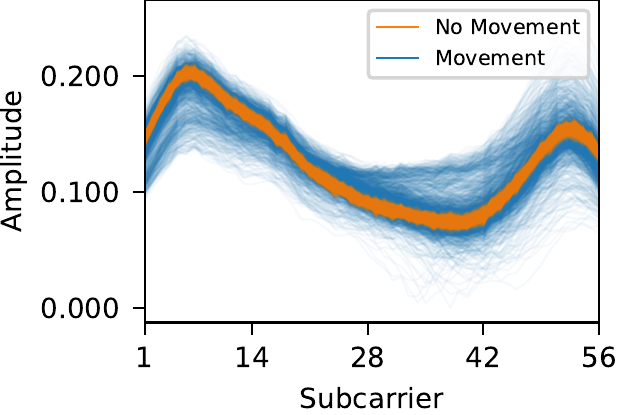}
\label{fig:attacker_processing__a}
} %
\subfloat[]{\includegraphics[width=0.49\linewidth]{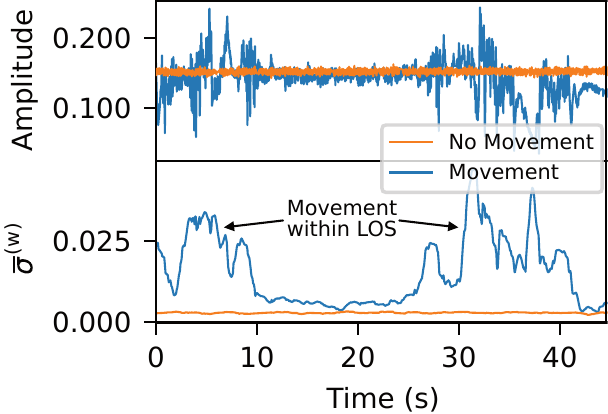}
\label{fig:attacker_processing__b}
}
\caption{(a)~Ensembles of raw CSI amplitudes over subcarriers from a standard \mbox{Wi-Fi} connection without and with movement in the propagation environment. (b)~Raw CSI time series of a single subcarrier and sliding window standard deviation with window length \SI{1}{\s} with and without movement.}
\label{fig:attacker_processing}
\end{figure}

The basic rationale for adversarial wireless sensing is that the wireless channel is coupled to environmental changes such as human motion. However, the exact relation between the observed channel and the environment is non-trivial, \ie, the channel response cannot be directly used to make elaborate claims on the physical state of the environment. Instead, several preprocessing steps shall be used in order to accomplish tasks such as motion detection. 
Here, we follow the attack strategy outlined in~\cite{zhu.2020}. Similar to \cite{banerjeeViolating.2014, WangLowHumanEffort.2018}, the authors describe an approach based on a sliding-window standard deviation over a time series of CSI magnitudes. The intuition behind this approach is rooted in the observation that a static channel response indicates a steady environment whereas a dynamic (time-varying) channel response indicates a varying environment. To illustrate this, we plot a set of \mbox{Wi-Fi} CSI values with and without movement in Fig.~\ref{fig:attacker_processing__a}.

By eavesdropping \mbox{Wi-Fi} packets at time $t$ from an anchor, Eve obtains a time series of complex-valued estimates of the wireless channel between Eve and the anchor node. At each time instant, Eve gathers a multitude of parallel channel estimates due to diversity from ($i$) the OFDM subcarriers and ($ii$) the spatial channels for MIMO transmissions. Therefore, Eve can leverage the combined measurements
\begin{align}
    \bm{h}(t)&=\left(\mathrm{vec}\left(\bm{H}(1,t)\right)^T,\dots,\mathrm{vec}\left(\bm{H}(K,t)\right)^T\right)^T,
\end{align}
to obtain insights on the victim's surroundings. In the following, we denote the $n^{\textrm{th}}$ element of such a measurement vector as $\bm{h}_{n}(t)$. 
Due to lacking synchronization, the signal phase of \mbox{Wi-Fi} channel estimates is subject to severe measurement imperfections and thus is not reliable. Therefore, we discard the phase and use the absolute value of $h_{n}(t)=\left|\bm{h}_{n}(t)\right|$. %
Then, Eve uses a windowing technique with window size~$w=$~\SI{1}{s} to calculate a moving standard deviation
\begin{align}
    \sigma^{(w)}_{n}(t) &=\underset{\tau\in \{t-w+1,...,t\}}{\mathbb{S}\mathrm{td}} \left(h_{n}(\tau) \right).%
\end{align}
To illustrate this step, Fig.~\ref{fig:attacker_processing__b} depicts a single \mbox{Wi-Fi} subcarrier over time~(top) and its corresponding sliding-window standard deviation~(bottom). It can clearly be seen that the latter quantifies motion-induced temporal variations.

Finally, to take advantage of the available diversity domains~\cite{banerjeeViolating.2014}, we average over $N_{\mathrm{R}} N_{\mathrm{T}}$~spatial channels and $K$ subcarriers, \ie, 
\begin{align}
    \bar{\sigma}^{(w)}(t) = \frac{1}{KN_{\mathrm{R}}N_{\mathrm{T}}}\sum_{n=1}^{KN_{\mathrm{R}}N_{\mathrm{T}}} \sigma^{(w)}_{n}(t).
\end{align}
In the remainder of this paper, we also refer to this averaged sliding-window standard deviation, $\bar{\sigma}^{(w)}(t)$, as \textit{adversarial observation}.

To decide whether $\bar{\sigma}^{(w)}(t)$ indicates motion, \cite{zhu.2020} utilizes a threshold-based detection, \ie, motion is detected if \mbox{$\bar{\sigma}^{(w)}(t)>u$}, where the threshold~$u$ is derived from a long-term reference measurement~$\bar{\sigma}^{(w)}_{\mathrm{ref}}(t)$ as
\begin{equation}
    u = \mathrm{median}_t\left\{\bar{\sigma}^{(w)}_{\mathrm{ref}}(t)\right\} + C \cdot \mathrm{MAD}_t\left\{\bar{\sigma}^{(w)}_{\mathrm{ref}}(t)\right\},
    \label{eq:threshold_motion}
\end{equation}
where $\mathrm{median}_t\left\{\cdot\right\}$ and $\mathrm{MAD}_t\left\{\cdot\right\}$ denote the median and median absolute deviation over $t$, respectively. According to the implementation details outlined in~\cite{zhu.2020}, we adopt the choice of the conservativeness factor $C=11$. It should be noted that alternative threshold definitions are possible such as, \eg, based on comparison of long-term and short-term sliding-window variances~\cite{banerjeeViolating.2014}.  

\subsection{Attack Setup}
\label{sec:router_setup}
Since our goal is to develop a countermeasure, we seek to strengthen the adversarial setup compared to~\cite{zhu.2020} where a smartphone-based eavesdropper with a single antenna was considered. %
Therefore, we utilize multi-antenna \mbox{Wi-Fi} routers (\mbox{TP-Link~N750}) for our experimental setup as both anchor nodes and eavesdroppers. These are equipped with $N_{\mathrm{R}}=N_{\mathrm{T}}=3$ antennas each and implement IEEE~802.11n~\mbox{Wi-Fi}~\cite{reference_routers}. In our experimental setup, the routers run an OpenWrt operating system and provide CSI data from the ath9k-based \mbox{Wi-Fi} chipset~\cite{xie_precise_2015} upon reception of a packet. The anchors transmit packets on a~\SI{20}{\MHz} wide channel at~\SI{5320}{\MHz} (\mbox{Wi-Fi} channel~$64$). Upon receipt of each packet, the eavesdropper obtains a complex-valued CSI vector, containing the channel estimations for each of the $9$~spatial channels and $56$ non-zero OFDM subcarriers. As in~\cite{zhu.2020}, out of the $56$ subcarriers, we select $K=28$ ones with the highest pair-wise correlation coefficient. Furthermore, while the smartphone implementation from~\cite{zhu.2020} reports $8$-$11$~packets/sec., our implementation is capable of approx.~$70$~packets/sec.

\begin{figure}
\centering
\includegraphics[width=0.7\linewidth]{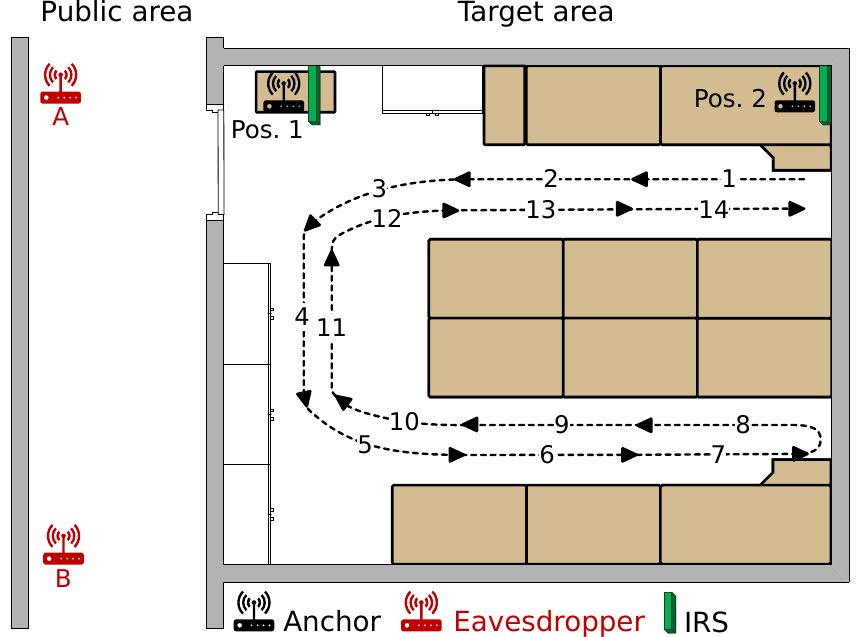}
\caption{The experimental setup with different positions of anchor nodes, eavesdroppers, and the motion path of a human.}
\label{fig:floorplan_motion_sensing_experiment}
\end{figure}

As the target environment, we consider an ordinary office within our institute's building of size approx.~\SI{7.5}~x~\SI{5.5}{\m}. A floor plan of the room (target area) is shown in Fig.~\ref{fig:floorplan_motion_sensing_experiment}. Here, at a publicly accessible area on the outside of the target area, the eavesdropper receives \mbox{Wi-Fi} packets from the anchors.

\subsection{Results}
\label{sec:attack_results}

In a real-world scenario, Eve would initially perform a long-term reference measurement to determine the motion-detection threshold $u$ using Eq.~\eqref{eq:threshold_motion}, assuming that the environment conditions remain steady over a long period of time~\cite{zhu.2020}. In our experiments, we grant the eavesdropper a \SI{3}{\min.} reference measurement under optimal conditions, \ie, the target area is guaranteed to be without motion and completely steady. %

\begin{figure}[!t]
\centering
\subfloat[Anchor $1$, Eve $A$]{\includegraphics[width=0.49\linewidth]{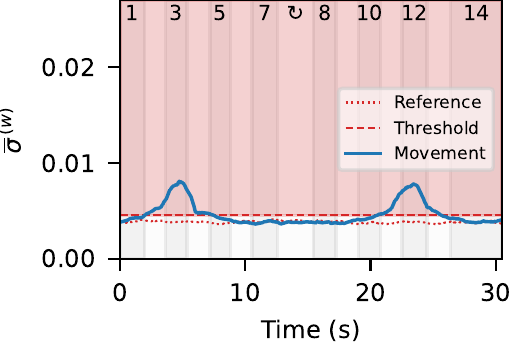}%
\label{fig:motion_sensing_attack1__a}%
} \hfill
\subfloat[Anchor $1$, Eve $B$]{\includegraphics[width=0.49\linewidth]{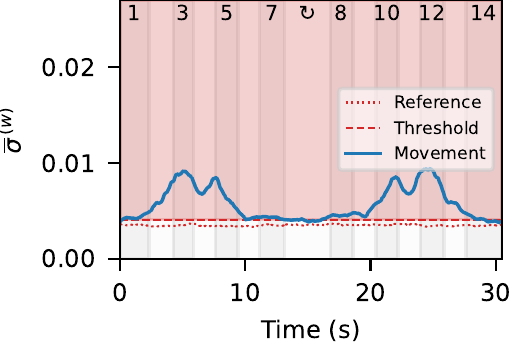}%
\label{fig:motion_sensing_attack1__b}%
}\\
\subfloat[Anchor $2$, Eve $A$]{\includegraphics[width=0.49\linewidth]{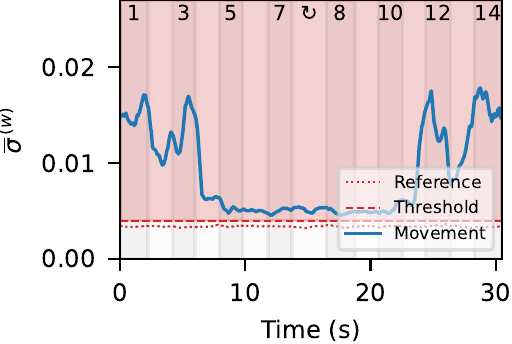}%
\label{fig:motion_sensing_attack1__c}%
} \hfill
\subfloat[Anchor $2$, Eve $B$]{\includegraphics[width=0.49\linewidth]{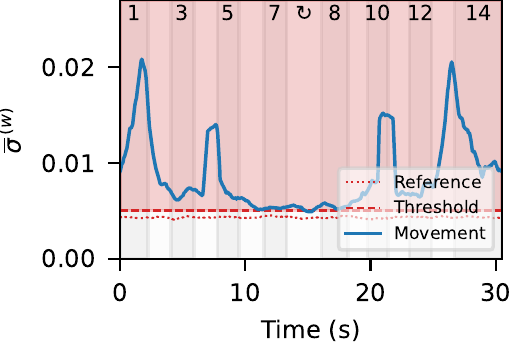}%
\label{fig:motion_sensing_attack1__d}%
}
\caption{Adversarial motion sensing results for eavesdropper positions~$A$ and $B$, receiving signals from anchor~$1$ and $2$ with a person walking along a pre-defined path (see the numbering in Fig.~\ref{fig:floorplan_motion_sensing_experiment}). We plot the channel sliding-window standard deviation observed by the adversary with and without movement in the target area.
}
\label{fig:motion_sensing_attack1}
\end{figure}

Subsequently, we proceed with a measurement where a person walks along a pre-defined path through the target area. The exact course is indicated by the numbered arrows in Fig.~\ref{fig:floorplan_motion_sensing_experiment}. The corresponding measurement results are shown in Fig.~\ref{fig:motion_sensing_attack1} where we plot the reference measurement and the derived threshold as well as the eavesdropper's observation during human motion.

Fig.~\ref{fig:motion_sensing_attack1__a} and~\ref{fig:motion_sensing_attack1__b} show the respective adversary observations by eavesdropping anchor~$1$ from positions~$A$ and~$B$ as depicted in Fig.~\ref{fig:floorplan_motion_sensing_experiment}. For eavesdropper position~$A$, we can see from the motion-trigger region (shaded red) that motion is successfully detected when the person is in proximity to the anchor device  (note the indication of the person's position at the top of each figure). For eavesdropper position~$B$, motion is detected with greater spatial extent, \ie, not only close to the anchor but also within the LOS of the anchor and the adversary. 

Firstly, this result confirms the attack scenario and the effectiveness of our implementation. Another key observation is the location-dependent attack performance: The motion detection works best when the person walking in the target environment is within or close to the LOS between the anchor and the eavesdropper. This observation is in accordance with the model from Section~\ref{sec:background-wsensing}: The strong LOS signal component and very direct NLOS signal components typically contribute most to the channel response. Thus, upon disturbance of these paths, the eavesdropper observes stronger variations in the channel response. This also lines up with the results in Fig.~\ref{fig:motion_sensing_attack1__c} and~\ref{fig:motion_sensing_attack1__d}, showing the observations from eavesdropping anchor~$2$. The LOS between the anchor and the eavesdropper now covers a larger area in the target environment and is crossed by the walking person several times. As depicted by the identical scaling of the $y$-axis for each figure, it can be clearly observed that the signals from anchor $2$ are more sensitive for human motion. Thus, this time the adversary successfully detects motion for the entire walk along the path.

\section{IRShield: Countermeasure Outline}\label{sec:irshield}
In this section, we describe our IRS-based countermeasure and provide further details about the IRS prototype used for our experiments. We then elaborate on the IRS-based channel obfuscation algorithm.%

\subsection{IRS-based Channel Obfuscation}
\label{sec:intro_irs_obfuscation}

As outlined in Section~\ref{sec:background_irs}, an IRS makes a portion of the wireless channel between a transmitter and a receiver adjustable. This technological novelty is our key building block to find an appropriate countermeasure against adversarial wireless physical-layer sensing: We introduce a randomly, time-varying IRS into the target area and thereby add randomness to the eavesdropper's channel observation to hamper detection of human motion. 

Crucially, the IRS is a powerful yet simple tool that the defender can use to invalidate the assumption underlying the attack: While variations of the channel previously allowed to conclude on environmental variations and human motion, the IRS now introduces artificial variations, effectively obfuscating the mapping of channel variations to motion in the environment. In order to engineer a countermeasure from this, the defender can address either the eavesdropper's threshold finding step or the actual motion sensing. From these two approaches, we can already formulate possible goals on desired effects of the countermeasure: ($i$)~Make the adversary pick an overly high threshold such that environmental variation does not trigger detection. ($ii$)~Let the adversary observe a strongly varying wireless channel such that the effect of human motion cannot be distinguished well. Additionally, since the adversary is passive, the defender is assumed to be unaware of the attack. Thus, we conclude that ($iii$)~the defense needs to operate continuously.

A question yet to be answered is how the IRS needs to be configured to effectively hamper the adversarial detection. Recall from Section~\ref{sec:attack} that the attack is based on a sliding-window standard deviation of channel responses where temporal fluctuations in the strength of channel variation triggers detection. Therefore, to mimic the effect of human motion, we conclude that the strength of IRS-induced channel variation should exhibit randomized temporal changes. %

\subsection{Experimental Setup}
\label{sec:irs_experimental_setup}

\begin{figure}
\centering
\includegraphics[width=1.0\linewidth]{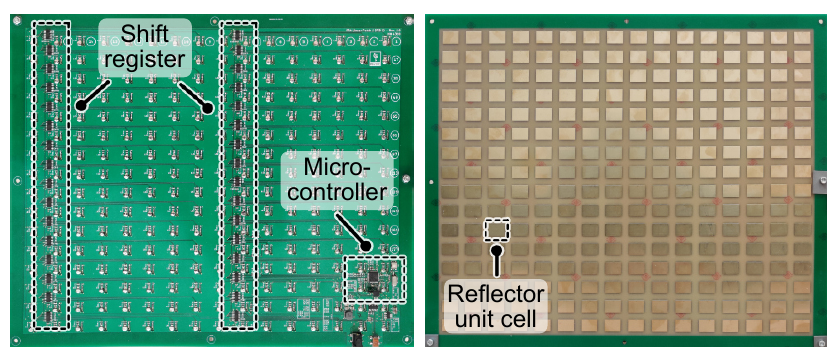}
\caption{Intelligent reflecting surface prototype (\SI{43}~$\times$~\SI{35}{\cm}). Back side with control circuitry (left) and front side with reflecting elements (right).}
\label{fig:irs_photo}
\end{figure}

For a first exploration of IRS-based channel obfuscation, we employ a prototypical IRS (see Fig.~\ref{fig:irs_photo}) with $256$~binary-phase tunable identical unit cell elements arranged in a $16 \times 16$~array on a \SI{43}~$\times$~\SI{35}{\cm} standard FR4 PCB substrate. The unit cell~\cite{heinrichs.2020} consists of a rectangular reflector patch whose RF reflection coefficient can be electronically switched by means of a PIN~diode to add a parasitic element to the reflector. By switching a bias voltage to the diode either on or off, the reflection coefficient of each IRS element can be individually switched between values represented by a `0'~state (off) and a `1'~state (on). The corresponding voltages are generated by means of cascaded shift registers, digitally controlled using a microcontroller. Leveraging a standard USB connection for serial communication, the surface's RF reflection behavior is elegantly programmed from a host computer via \SI{256}{\bit} words. 
The IRS used in our experiments is optimized to achieve a \SI{180}{\degree} phase shift at around~\SI{5.35}{\GHz} in the reflected wave when switching between the `0' and `1'~states, \ie, $r_m \in \{-1, 1\}$ in Eq.~\eqref{eq:channel_multipath_irs}. %

\subsection{IRS Configuration Strategy}
\label{sec:irs_cfg_strategy}

We now proceed with the design of a plug-and-play configuration strategy to achieve IRS-based channel obfuscation in the field. Within the eavesdropper's channel response, \cf~Eq.~\eqref{eq:channel_multipath_irs}, the contribution of the IRS is given as 
\begin{align}
    \bm{H}_{\textrm{IRS}}(k,t)=\sum_{m=1}^{M}r_m(t)\bm{G'}_m(k,t).
    \label{eq:H_irs_attacker}
\end{align}
Here we can see that the $M$~individual IRS reflection coefficients $r_m(t)$ adjust the multipath components $\bm{G'}_m(k,t)$. The key observation to be made is that the signal amplitude from superposition of the multipath components in Eq.~\eqref{eq:H_irs_attacker} depends on the IRS configuration. Thus, the amplitude of the eavesdropper's observation, \cf~Eq.~\eqref{eq:channel_multipath_irs}, will likewise depend on the IRS. %
It should be noted that the mapping of specific IRS configurations to a particular channel response is non-trivial. As we here consider a binary-phase tunable IRS with $256$~elements, there are $2^{256}$ possible realizations of $\bm{H}_{\textrm{IRS}}(k,t)$. In view of the large design space and the unknown channels to the eavesdropper $\bm{G'}_m(k,t)$ (recall that Eve is completely passive), it is not possible to distill or predict tailored IRS configurations.

Therefore, we pursue the following probabilistic approach where we gradually change the IRS configuration. Given an arbitrary initial set $\{r_m\}$, in the first step, we start by inverting a small amount of randomly chosen elements, \eg, \SI{5}{\percent} of all $256$~elements. In the next step, we invert all $256$~elements and go back to the beginning to repeat the procedure. Thereby, the IRS configuration will change gradually but random and similarly will the amplitude of the resulting IRS signal, thus yielding smooth amplitude gradients. %
The intuition behind inverting the IRS configuration is the resulting \SI{180}{\degree} phase shift of $\bm{H}_{\textrm{IRS}}(k,t)$, \ie, when $r_m \in \{-1, 1\}$, we can factor out $-1$ in Eq.~\eqref{eq:H_irs_attacker} such that $\bm{H}^{(0)}_{\textrm{IRS}}(k,t) = -\bm{H}^{(1)}_{\textrm{IRS}}(k,t)$. Thus, in combination with the non-IRS propagation in~Eq.~\eqref{eq:channel_multipath_irs}, a surface inversion will yield strong amplitude changes in the wireless channel.

We implement the outlined configuration strategy as further described by Algorithm~\ref{alg:surface_configuration}. The algorithm leverages a state machine to switch back and forth between changing a small number of elements and inverting all elements. Additionally, the algorithm timing is randomized by skipping the state transition with a probability $P_{\mathrm{hold}}$, with which the IRS configuration remains unchanged. In this way, gaps in the IRS operation cause drops in channel variation which enhance the median absolute deviation in Eq.~\eqref{eq:threshold_motion}.

\begin{figure}
\centering
\includegraphics[width=0.97\linewidth]{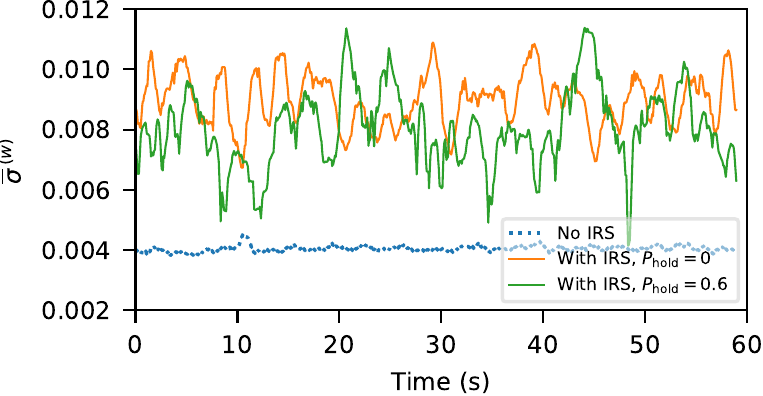}
\caption{Illustration of the effect of IRShield on the adversarial channel observation. We plot the channel sliding-window standard deviation with and without IRShield.}
\label{fig:defense_intuition}
\end{figure}

We test the effectiveness of the algorithm and place an IRS close, \ie, \SI{30}{\cm}, to the anchor position $2$, and record channel observations of the eavesdropper at position $A$ without any motion in the target environment. We first conduct a reference measurement with the IRS being disabled. Then, we use Algorithm~1 to configure the IRS at an update rate of $20$~configurations/sec, and record channel measurements with $P_{\mathrm{hold}} = 0$ and $P_{\mathrm{hold}} = 0.6$. We plot the results as time series in Fig.~\ref{fig:defense_intuition}. With the IRS being inactive, the channel remains static as expected. However, as can be seen from the plot, the IRS operation succeeds to produce strong time variations. Furthermore, the effect of $P_{\mathrm{hold}}$ is evident from pronounced excursions with low channel variation. As our results show, the IRS configuration algorithm meets the requirement to generate channel variations of varying strength. We provide additional insight and experimental results on the algorithm parametrization in Appendix~\ref{sec:algorithm_parameters}. 

\begin{algorithm}[ht]
\small
\DontPrintSemicolon
\SetAlgoLined
 random IRS configuration $\textrm{cfg}[M] \in \{0, 1\}$;\;
 progression rate $R=0.05$;\;
 hold probability $P_{\mathrm{hold}} = 0.6$;\;
 nextState = `RAND';\;
 \While{True}{
    \uIf{getRandom(0,1) $< P_{\mathrm{hold}}$}{
         \Comment*[r]{Remain in current state}
    }
    \Else{
        currState = nextState;\;
        \uIf{currState == `RAND'}{
            $T \gets \ceil*{R \cdot M}$ random unique IRS elements;\;
            \ForEach{$m \in T$}{
                $\textrm{cfg}[m] = \textrm{cfg}[m] \oplus 1$; \Comment*[r]{Flip selected IRS elements}
            }
            nextState = `FLIP';\;
        }
        \ElseIf{currState == `FLIP'}{
            $\textrm{cfg} = \textrm{cfg} \oplus 1$; \Comment*[r]{Flip all elements}
            nextState = `RAND';\;
        }
    }
    write $\textrm{cfg}$ to IRS;\;
 }
 \caption{IRS configurations for IRShield.}
 \label{alg:surface_configuration}
\end{algorithm}

\section{Results}\label{sec:results}

In this section, we experimentally investigate how IRShield %
affects adversarial motion detection. Further, we examine the effect of size, distance, and orientation of the IRS.%

\subsection{Human Motion Sensing}
\begin{figure*}[t!]
\centering
\subfloat[Anchor $1$, Eve $A$]{\includegraphics[width=0.49\linewidth]{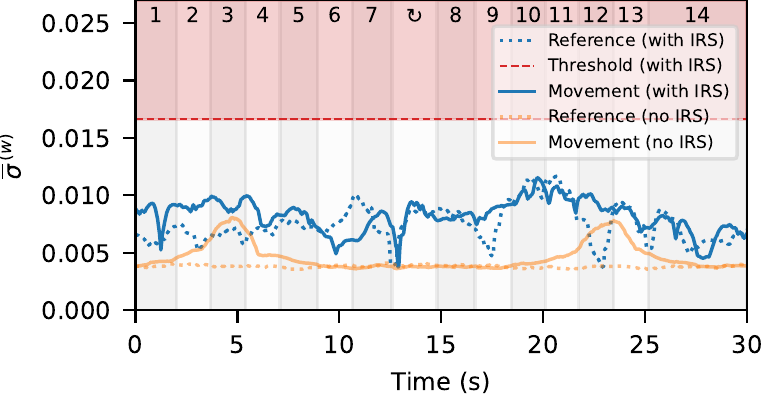}%
\label{fig:motion_sensing_attack1_wIRS__a}%
} \hfill
\subfloat[Anchor $1$, Eve $B$]{\includegraphics[width=0.49\linewidth]{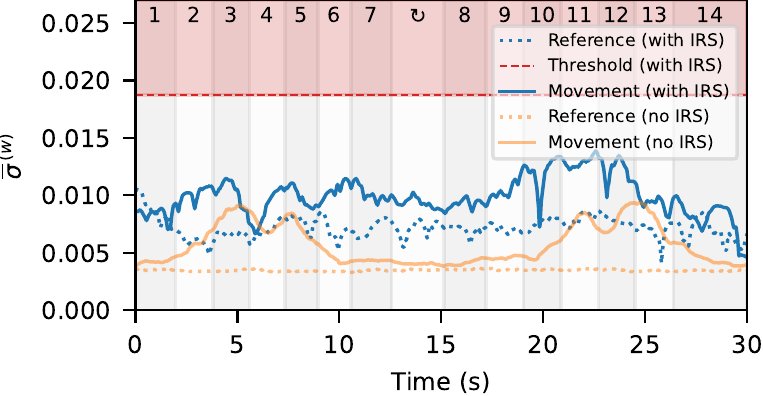}%
\label{fig:motion_sensing_attack1_wIRS__b}%
}\\
\subfloat[Anchor $2$, Eve $A$]{\includegraphics[width=0.49\linewidth]{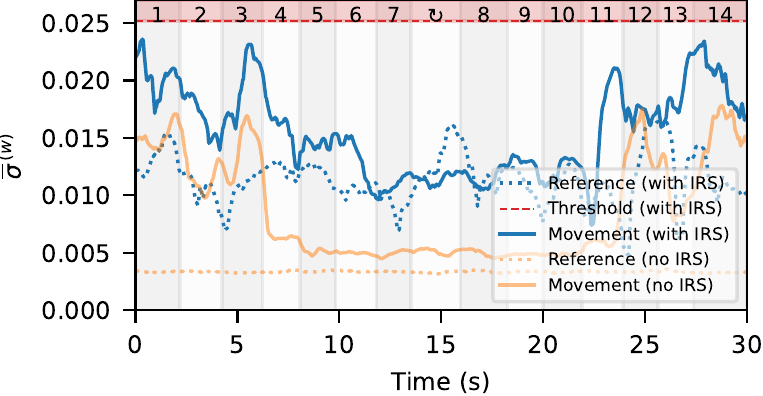}%
\label{fig:motion_sensing_attack1_wIRS__c}%
} \hfill
\subfloat[Anchor $2$, Eve $A$]{\includegraphics[width=0.49\linewidth]{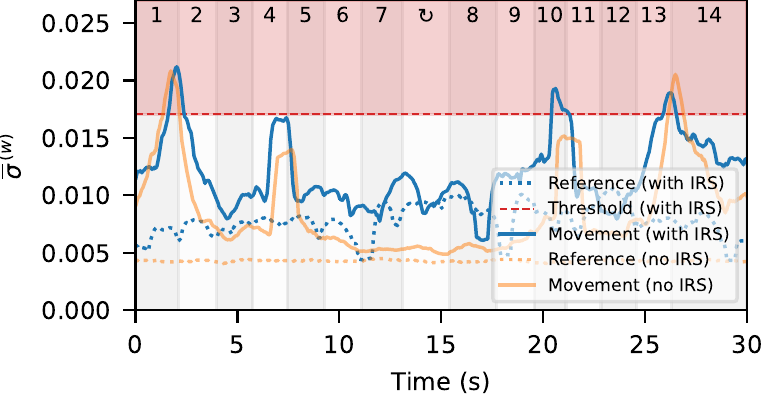}%
\label{fig:motion_sensing_attack1_wIRS__d}%
}
\caption{Adversarial motion sensing results with IRShield active. We plot the channel sliding-window standard deviation observed by the eavesdropper during the reference measurement without motion and with a person walking along a pre-defined path (see the numbering in Fig.~\ref{fig:floorplan_motion_sensing_experiment}). Also, we indicate the motion-detection threshold according to~\cite{zhu.2020} with $C=11$. For comparison, we also plot the results when IRShield is not active, \cf~Fig.~\ref{fig:motion_sensing_attack1}.}
\label{fig:motion_sensing_attack1_wIRS}
\end{figure*}

Putting the IRShield channel obfuscation to test, we next investigate its impact on the detection of motion. Therefore, we repeat the human-motion sensing experiments from Section~\ref{sec:attack}, this time with IRShield in place. Here, we again use the experimental setup illustrated in Fig.~\ref{fig:floorplan_motion_sensing_experiment}, indicating a walking path and positions of the anchor device, the IRS, and the eavesdropper. For the IRS, we use the setup described in Section~\ref{sec:irs_experimental_setup}. Initially, the eavesdropper gathers a \SI{3}{min.} long reference measurement without any motion to determine a motion-detection threshold, however, observing not only the steady environment but also the time-varying IRS. Next, with the IRS still active, the eavesdropper records channel measurements while a person walks along the pre-defined path.

In Fig.~\ref{fig:motion_sensing_attack1_wIRS}, we plot the results from four combinations of anchor-eavesdropper positions as time series, showing the reference measurement without motion, the resulting detection threshold according to~\cite{zhu.2020}, and the measurement with motion. As the baseline, we additionally plot the results without the IRS countermeasure, as presented in~Fig.~\ref{fig:motion_sensing_attack1}. When comparing results without IRShield, the first thing to note is that the adversary fails to detect motion in almost every case. This is because the adversary's reference measurement is degraded by the defender's IRS-induced channel variation, resulting in an excessively high threshold value. For anchor position~$1$, Fig.~\ref{fig:motion_sensing_attack1_wIRS__a} and~\ref{fig:motion_sensing_attack1_wIRS__b} show the results when Eve is located at positions~$A$ and~$B$, respectively. In this case, the reference observation without motion strongly overlaps with the curve corresponding to motion. Thus, we conclude that our countermeasure renders adversarial wireless motion detection infeasible -- even when adjusting the threshold. However, the situation is slightly different when considering less favorable anchor placement, \eg, anchor position~$2$ with a large LOS, for which we show the results in Fig.~\ref{fig:motion_sensing_attack1_wIRS__c} and~\ref{fig:motion_sensing_attack1_wIRS__d}. In this case, the IRS operation affects the adversary's threshold finding such that motion is poorly detected. However, it appears that LOS crossings may in principle still be detectable if the adversary is aware of the channel obfuscation and accordingly adjusts the threshold. Thus, we conclude that the effect of IRShield, just like the attack itself, is location-dependent. Therefore, we next systematically study spatial dependencies.

\subsection{Systematic Coverage Tests}
\label{sec:heatmap}
As we have outlined previously, due to the varying strength of different RF propagation paths, the adversarial motion sensing is subject to location-dependent performance and this likewise appears to be the case for our IRS-based countermeasure. Therefore, we now seek to answer \textit{where} in the target area motion can be detected. 

For this, we test a total of $20$~positions within the target area. For each of the positions, distributed on a uniform grid throughout the target area (see Fig.~\ref{fig:systematic_experiment_setup} in Appendix~\ref{sec:coverage_setup}), we place a rotating reflector as a point source of repeatable and RF-relevant motion. The reflector is a \SI{50}~$\times$~\SI{50}{\cm} aluminum sheet, mounted on a motor to rotate at~approx.~\SI{20}{rpm}. After an initial \SI{3}{min.} long reference measurement without any motion in the target area, we record the adversary's channel observations from a particular anchor device with the rotating reflector at each of the $20$~positions. We repeat this procedure with and without IRShield being active. To quantify the adversary's success, we use the motion-detection rate using a threshold found from the initial reference measurement according to \cite{zhu.2020}. The detection rate is given as the fraction of total motion observations that lie above the threshold.

\begin{figure}[t]
\centering
\subfloat[Anchor $1$, Eve $A$, without IRS]{\includegraphics[width=0.49\linewidth]{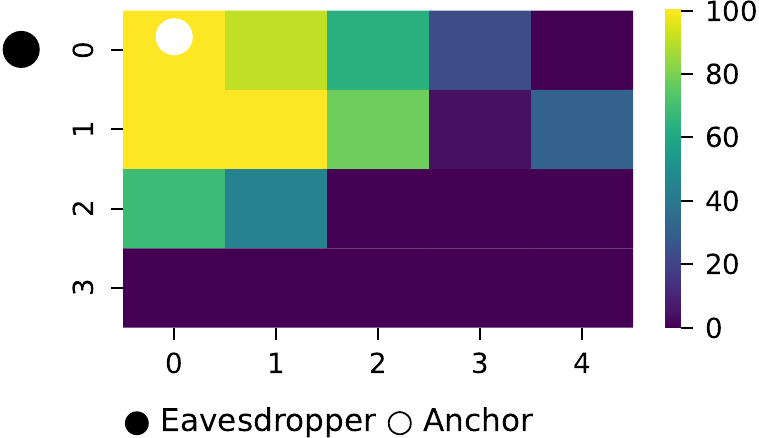}%
\label{fig:heatmaps_NLOS_alexa__a}%
} \hfill
\subfloat[Anchor $1$, Eve $A$, with IRS]{\includegraphics[width=0.49\linewidth]{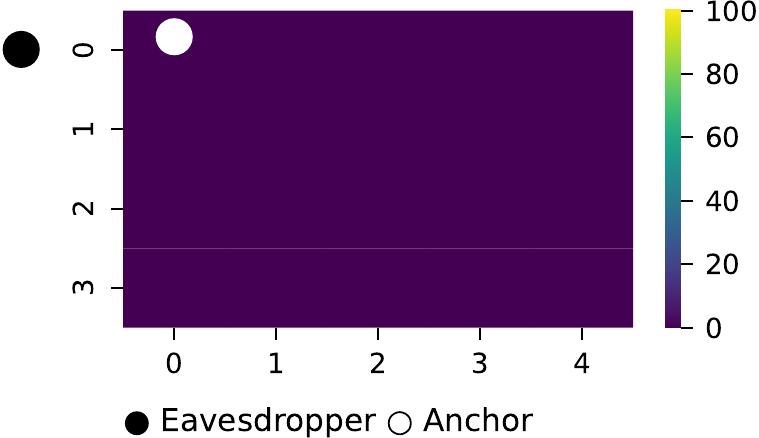}%
\label{fig:heatmaps_NLOS_alexa__b}%
}\\
\subfloat[Anchor $1$, Eve $B$, without IRS]{\includegraphics[width=0.49\linewidth]{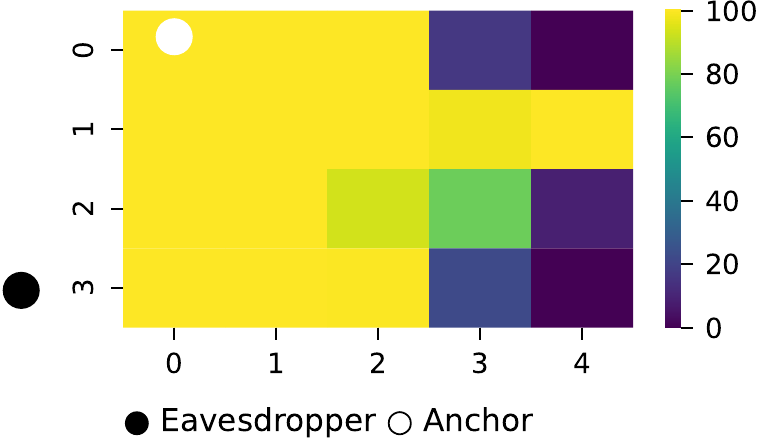}%
\label{fig:heatmaps_NLOS_alexa__c}%
} \hfill
\subfloat[Anchor $1$, Eve $B$, with IRS]{\includegraphics[width=0.49\linewidth]{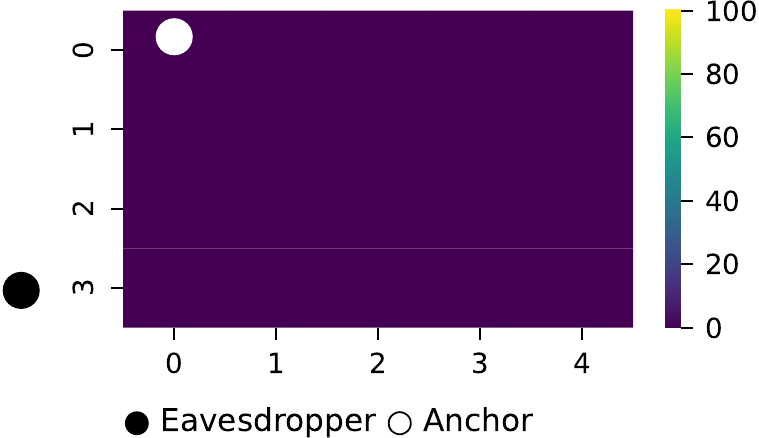}%
\label{fig:heatmaps_NLOS_alexa__d}%
}
\caption{Spatial distribution of detection rates for anchor position $1$ with motion-detection threshold according to~\cite{zhu.2020} with $C=11$.} 
\label{fig:heatmaps_NLOS_alexa}
\end{figure}

For anchor position~$1$ and two eavesdropper positions, we show the spatial distribution of detection rates without (Figs.~\ref{fig:heatmaps_NLOS_alexa__a} and \ref{fig:heatmaps_NLOS_alexa__c}) and with (Figs.~\ref{fig:heatmaps_NLOS_alexa__b} and \ref{fig:heatmaps_NLOS_alexa__d}) the IRS being active. Without the countermeasure, as expected, we can observe the highest detection rates within and around the LOS. Far off the LOS, the detection rate drops down to \SI{0}{\percent}. This effect is best visible in Fig.~\ref{fig:heatmaps_NLOS_alexa__a}: Due to the adversary's strong LOS signal, the sensitivity for weaker (environment-dependent) multipath signals is reduced. In contrast, when the IRS is active, the adversarial detection is completely suspended as evident from the \SI{0}{\percent} detection rate at all positions.

\begin{figure}[t]
\centering
\subfloat[Anchor $2$, Eve $A$, without IRS]{\includegraphics[width=0.49\linewidth]{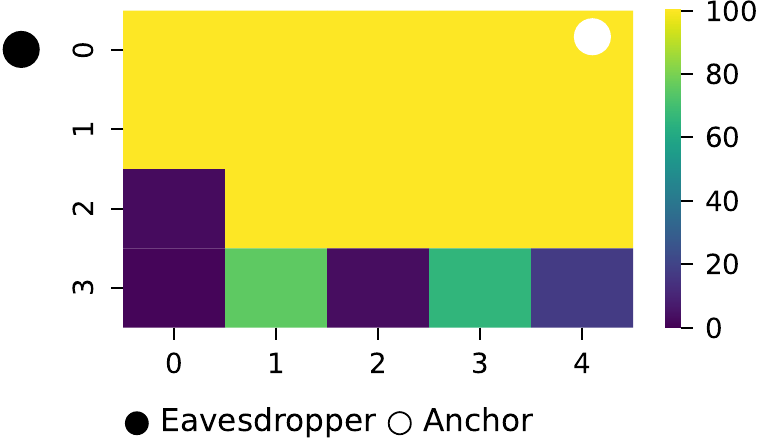}%
\label{fig:heatmaps_LOS_alexa__a}%
} \hfill
\subfloat[Anchor $2$, Eve $A$, with IRS]{\includegraphics[width=0.49\linewidth]{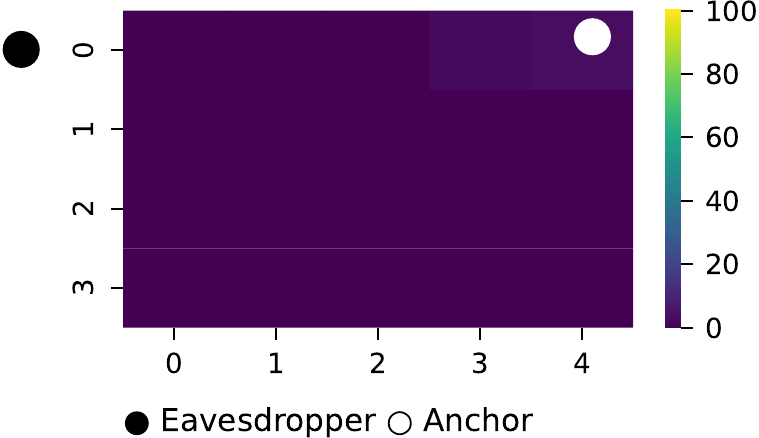}%
\label{fig:heatmaps_LOS_alexa__b}%
}\\
\subfloat[Anchor $2$, Eve $B$, without IRS]{\includegraphics[width=0.49\linewidth]{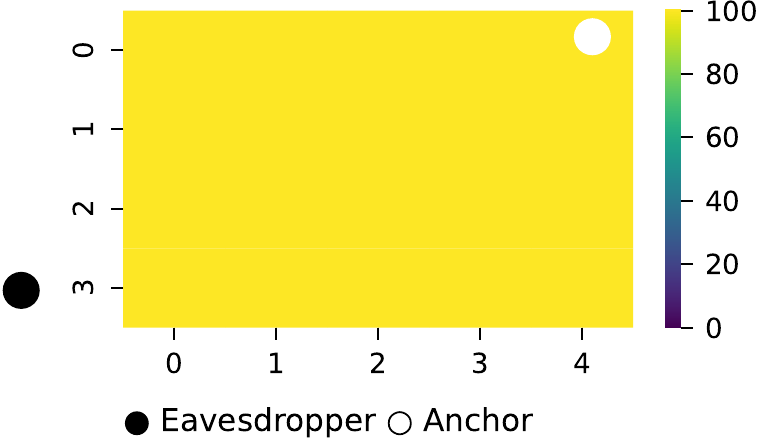}%
\label{fig:heatmaps_LOS_alexa__c}%
} \hfill
\subfloat[Anchor $2$, Eve $B$, with IRS]{\includegraphics[width=0.49\linewidth]{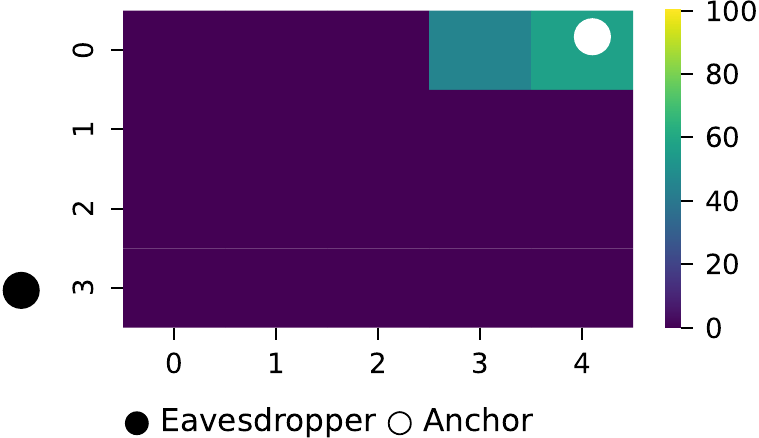}%
\label{fig:heatmaps_LOS_alexa__d}%
}
\caption{Spatial distribution of detection rates for anchor position $2$ with motion-detection threshold according to~\cite{zhu.2020} with $C=11$.} 
\label{fig:heatmaps_LOS_alexa}
\end{figure}

For the same eavesdropper positions and for anchor position~$2$ (large LOS), we show the results in Fig.~\ref{fig:heatmaps_LOS_alexa}. While Fig.~\ref{fig:heatmaps_LOS_alexa__a} again highlights the attack's LOS-dependent coverage, the diagonal LOS through the target area in Fig.~\ref{fig:heatmaps_LOS_alexa__c} allows the eavesdropper to obtain \SI{100}{\percent} detection rate on each position. Even for this challenging scenario, our countermeasure succeeds to reduce the detection rate to \SI{0}{\percent} on all positions (Fig.~\ref{fig:heatmaps_LOS_alexa__b}) and $18$ out of $20$ positions (Fig.~\ref{fig:heatmaps_LOS_alexa__d}). %

\Paragraph{Threshold adjustments}
\begin{table*}[!htb]
\small
\centering
\caption{Average detection rates for various thresholds, and placements of anchor and eavesdropper.}
\label{tab:detection_rates_text}
\begin{tabular}{@{}rcccc|cccc|cccc|cccc@{}}
\toprule
& \multicolumn{4}{c}{\textbf{Anchor 1, eavesdropper A}} & \multicolumn{4}{c}{\textbf{Anchor 1, eavesdropper B}} & \multicolumn{4}{c}{\textbf{Anchor 2, eavesdropper A}} & \multicolumn{4}{c}{\textbf{Anchor 2, eavesdropper B}}\\ \cmidrule(lr){2-5} \cmidrule(lr){6-9} \cmidrule(lr){10-13} \cmidrule(lr){14-17}
& \multicolumn{2}{c}{\textbf{No IRS}} & \multicolumn{2}{c}{\textbf{With IRS}} & \multicolumn{2}{c}{\textbf{No IRS}} & \multicolumn{2}{c}{\textbf{With IRS}} & \multicolumn{2}{c}{\textbf{No IRS}} & \multicolumn{2}{c}{\textbf{With IRS}} & \multicolumn{2}{c}{\textbf{No IRS}} & \multicolumn{2}{c}{\textbf{With IRS}}\\ \cmidrule(lr){2-3} \cmidrule(lr){4-5} \cmidrule(lr){6-7} \cmidrule(lr){8-9} \cmidrule(lr){10-11} \cmidrule(lr){12-13} \cmidrule(lr){14-15} \cmidrule(lr){16-17} 
& TPR & FPR & TPR & FPR & TPR & FPR & TPR & FPR & TPR & FPR & TPR & FPR & TPR & FPR & TPR & FPR\\
\midrule
$\mathrm{max}_t\left\{\cdot\right\}$ & 0.64 & 0.00 & 0.03 & 0.00 & 0.92 & 0.00 & 0.08 & 0.00 & 0.94 & 0.00 & 0.28 & 0.00 & 1.00 & 0.00 & 0.39 & 0.00\\
$C=1$ & 0.91 & 0.25 & 0.42 & 0.26 & 1.00 & 0.26 & 0.62 & 0.28 & 0.99 & 0.27 & 0.62 & 0.27 & 1.00 & 0.25 & 0.86 & 0.23\\
$C=3$ & 0.75 & 0.02 & 0.11 & 0.04 & 0.98 & 0.04 & 0.23 & 0.04 & 0.98 & 0.04 & 0.41 & 0.04 & 1.00 & 0.03 & 0.55 & 0.02\\
$C=5$ & 0.61 & 0.00 & 0.03 & 0.00 & 0.94 & 0.00 & 0.04 & 0.00 & 0.96 & 0.00 & 0.26 & 0.00 & 1.00 & 0.00 & 0.30 & 0.00\\
$C=7$ & 0.51 & 0.00 & 0.00 & 0.00 & 0.88 & 0.00 & 0.00 & 0.00 & 0.91 & 0.00 & 0.16 & 0.00 & 1.00 & 0.00 & 0.15 & 0.00\\
$C=9$ & 0.42 & 0.00 & 0.00 & 0.00 & 0.82 & 0.00 & 0.00 & 0.00 & 0.83 & 0.00 & 0.06 & 0.00 & 1.00 & 0.00 & 0.08 & 0.00\\
$C=11$~\cite{zhu.2020} & 0.35 & 0.00 & 0.00 & 0.00 & 0.76 & 0.00 & 0.00 & 0.00 & 0.78 & 0.00 & 0.00 & 0.00 & 1.00 & 0.00 & 0.05 & 0.00\\
\bottomrule 
\end{tabular}
\end{table*}
We evaluated the specific threshold finding of the original attack as proposed by~\cite{zhu.2020}. However, choosing the threshold less conservative will slightly improve the detection rates. For each of the four combinations of anchor and eavesdropper positions, Table~\ref{tab:detection_rates_text} lists the average detection rates for varying motion-detection thresholds. For the state-of-the-art attack~\cite{zhu.2020}, IRShield lowered detection rates to~\SI{5}{\percent} or less, as can be seen from the last row, with $C=11$ in~Eq.~\eqref{eq:threshold_motion}, corresponding to Fig.~\ref{fig:heatmaps_NLOS_alexa} and Fig.~\ref{fig:heatmaps_LOS_alexa}. The first row, indicated by $\mathrm{max}_t\left\{\cdot\right\}$, gives the results with the maximum of the reference measurement being used as motion-detection threshold. Thus, the detection rates were obtained with the lowest threshold resulting in a false positive rate (FPR) of $0$ (at least for the observed reference duration). The respective heatmaps are shown in Appendix~\ref{sec:heatmaps_reduced_threshold} (Fig.~\ref{fig:heatmaps_NLOS_maxref} and Fig.~\ref{fig:heatmaps_LOS_maxref}). 

\begin{figure}
\centering
\centering
\subfloat[Eavesdropper position $A$]{\includegraphics[width=0.49\linewidth]{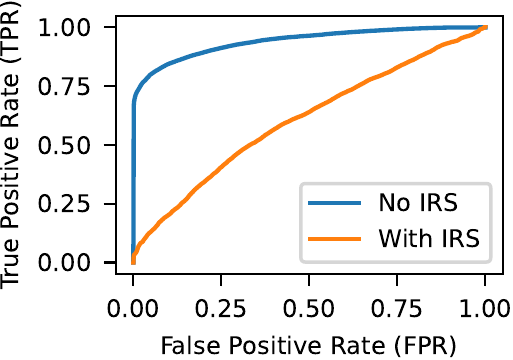}%
} \hfill
\subfloat[Eavesdropper position $B$]{\includegraphics[width=0.49\linewidth]{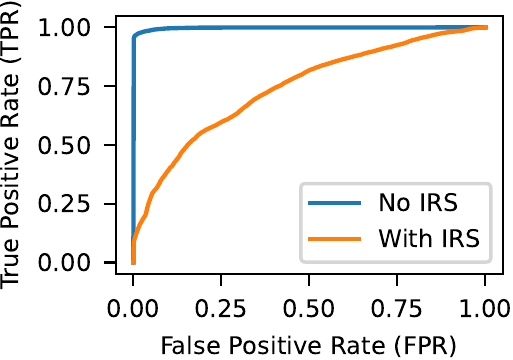}%
}
\caption{ROC curves with and without IRShield for anchor position $1$ and (a) eavesdropper position $A$, (b) eavesdropper position $B$.}
\label{fig:rocs}
\end{figure}

For anchor position~$2$, an adapted threshold would allow the adversary to restore some detection capabilities. Regarding the spatial aspects of this result, it can be seen from Fig.~\ref{fig:heatmaps_NLOS_maxref} and Fig.~\ref{fig:heatmaps_LOS_maxref} in Appendix~\ref{sec:heatmaps_reduced_threshold} that motion within the LOS and directly adjacent to the anchor is still reliably detected. This result is a vivid example for a limitation of IRS-based channel obfuscation: Due to the finite size of the IRS, it reflects a limited amount of signal power. Thus, it is challenging for the IRS to produce channel variations as strong as LOS crossings. On the other hand, for anchor position~$1$, the eavesdropper does not observe a signal with significant LOS coverage. This is reflected in Table~\ref{tab:detection_rates_text}, showing that the true positive rate (TPR) cannot be substantially increased without also increasing the FPR. To illustrate the separability of the eavesdropper's observations without motion and with motion, we plot the receiver operating curves~(ROCs) in Fig.~\ref{fig:rocs} for anchor position~$2$. The plots show the trade-off between the TPR and FPR, essentially visualizing all possible choices of detection thresholds. Especially for eavesdropper position~$A$, the detection approaches a random classifier. This result highlights that IRS-based channel obfuscation under certain conditions provides near-ideal channel randomization, thus thwarting adversarial motion detection.

\subsection{Impact of IRS}
Next, we seek to further characterize the effect of varying IRS parameters to provide a thorough guideline for practical deployments. Therefore, we change conditions such as size, distance, and orientation of the IRS w.r.t. an anchor device. For experimental simplicity, we test these conditions without considering human motion and leverage the strength of IRS-induced channel variation as a proxy for the effectiveness of channel obfuscation.

\Paragraph{Effective IRS size}
The IRS is passive w.r.t. RF signals, \ie, it only reflects but does not amplify signals. The physical size of the IRS is an important factor which limits the reflected signal power~\cite{wuSmartReconfigurableEnvironment2020}. Thus, the IRS size is one parameter that affects the effectiveness of IRShield. Studying this connection experimentally, we vary the effective IRS size (since the physical size of the IRS is fixed) by simply deactivating a certain number of IRS elements. Since IRShield is based on time-varying surface configurations, the deactivated (static) elements will not contribute to the channel variation. 

\begin{figure}
\centering
\includegraphics[width=0.97\linewidth]{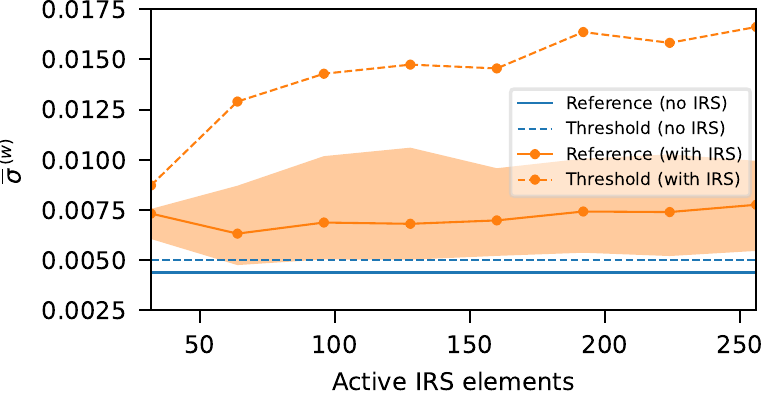}
\caption{Effect of varying number of active IRS elements. We plot the median and the $1^{\textrm{st}}$ and $99^{\textrm{th}}$ percentiles of the adversarial observations and the resulting motion-detection threshold according to~\cite{zhu.2020} with $C=11$.}
\label{fig:IRS_element_count}
\end{figure}

In our experiment, we increase the number of active IRS elements $M$ from~$32$ to~$256$ in steps of $32$ while we use Algorithm~\ref{alg:surface_configuration} to generate IRS configurations. The IRS elements to be added to the set of active elements are selected randomly. For each effective surface size, we then record \SI{2}{min.} of adversarial observations under channel obfuscation. We plot the median, the $1^{\textrm{st}}$ and $99^{\textrm{th}}$ percentiles of the channel variation together with the corresponding motion-detection threshold according to~\cite{zhu.2020} over the number of active IRS elements in Fig.~\ref{fig:IRS_element_count}. We can see that the median and the threshold grow with the effective IRS size. This confirms our hypothesis that an increasing number of varying IRS elements enhances channel obfuscation. However, since the adversary's motion-detection threshold is already substantially increased for $64$~active elements, we conclude that IRShield does not necessarily require large IRS deployments. %

\Paragraph{IRS distance}
Next, we investigate the connection between IRS-based channel obfuscation and the IRS distance to the anchor device. Signal propagation including an IRS is typically modeled by a multiplication of channels from the transmitter to the IRS and from the IRS to the receiver~\cite{wuSmartReconfigurableEnvironment2020}. Thus, due to the inherent distance-dependent path loss of wireless radio channels, the received power from the IRS will be inversely proportional to the inverse squared product of the distances to and from the IRS~\cite{ozdoganIntelligentReflectingSurfaces2020}. That is, the IRS works best near the transmitter or receiver (since one of the IRS distances is minimized). However, specific to our channel obfuscation application for, \eg, living spaces, it is reasonable to assume that the IRS is located rather close to both anchor devices and the receiver (the eavesdropper). 

\begin{figure}
\centering
\includegraphics[width=0.97\linewidth]{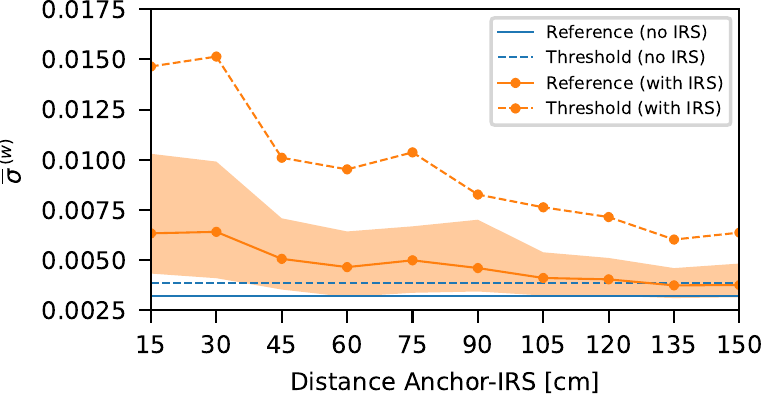}
\caption{Effect of varying distances of the IRS to the anchor device. We plot the median and the $1^{\textrm{st}}$ and $99^{\textrm{th}}$ percentiles of the adversarial observations and the resulting motion-detection threshold according to~\cite{zhu.2020} with $C=11$.}
\label{fig:distance_IRS_anchor}
\end{figure}

To assess this experimentally, we vary the distance of the IRS to anchor~$2$. For each distance, the eavesdropper observes the wireless channel from position~$A$ for \SI{3}{min.}. We plot the median, the $1^{\textrm{st}}$ and $99^{\textrm{th}}$ percentiles of the channel variation and the corresponding motion-detection threshold according to~\cite{zhu.2020} over IRS distances between \SI{15}~and~\SI{150}{\cm} in Fig.~\ref{fig:distance_IRS_anchor}. As expected, the channel obfuscation works best when the IRS is close to the anchor. Please note that the IRS generally can operate from higher distances as well, however, requiring optimal IRS configurations which the probabilistic Algorithm~\ref{alg:surface_configuration} is not designed for.

\begin{figure}
\centering
\includegraphics[width=0.9\linewidth]{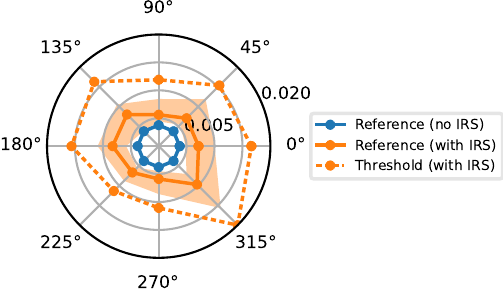}
\caption{Effect of IRShield on the adversarial reference measurement for varying IRS orientations. The IRS faces towards the eavesdropper at \SI{0}{\degree}. We plot the median and the $1^{\textrm{st}}$ and $99^{\textrm{th}}$ percentiles of the adversarial observations and the resulting motion-detection threshold according to~\cite{zhu.2020} with $C=11$.}
\label{fig:IRS_orientation}
\end{figure}

\Paragraph{IRS orientation}
Despite the distance, another relevant aspect of IRS placement is its orientation w.r.t. the eavesdropper. In the previous experiments, we placed the IRS facing towards the eavesdropper, \ie, the front of the IRS pointed towards the left side of the target area shown in Fig.~\ref{fig:floorplan_motion_sensing_experiment}. As this placement presumably is optimal for the defender, we additionally test less optimal orientations by rotating the IRS around an anchor device. For each IRS placement (at a constant distance of \SI{30}{\cm}), the adversary records channel measurements for \SI{1}{min.}. Fig.~\ref{fig:IRS_orientation} shows the median, the $1^{\textrm{st}}$ and $99^{\textrm{th}}$ percentiles of the channel variation in a polar plot, where the angle indicates the IRS placement. For \SI{0}{\degree}, the IRS faces towards the eavesdropper. As expected, we can see that the channel obfuscation works best when the IRS is roughly directed towards the eavesdropper. Thus, when the environment to be protected has limited public accessibility (\eg, a hallway on one side of the property such as in Fig.~\ref{fig:floorplan_motion_sensing_experiment}), the defender should point the IRS towards this direction. However, despite that, even in the worst case where the IRS points in the opposite direction of the eavesdropper, the channel obfuscation is still effective, albeit with reduced impact.

\Paragraph{Data throughput}
Another important question is whether IRShield affects the wireless data throughput. Previous countermeasures, \eg, \cite{zhu.2020} and \cite{wijewardena.2020}, affect the allocation of the transmission medium and therefore face a trade-off between the quality of service of wireless communication and reduction of adversarial sensing capabilities. To this end, we emphasize that our channel obfuscation approach is based on changing the transmission medium itself instead of its allocation. Therefore, we expect the channel obfuscation to have a marginal effect on the data throughput. %

We put this claim to test by measuring packet error rates~(PERs) of a \mbox{Wi-Fi} connection with and without the channel obfuscation being active. For this, we deploy two legitimate IEEE~802.11n \mbox{Wi-Fi} devices in the eavesdropper's target area and make one device transmit \SI{200000} packets while the other keeps track of the received packets. For the transmissions, we randomize the payload data and use a fixed modulation and coding scheme~(MCS)~\cite{mcs_index_website}. In this experiment, we use $2$~$\times$~$2$~MIMO devices which calls for using MCS~values~\mbox{$8$-$15$} (two spatial streams). We place the IRS next to the transmitter.

\begin{figure}
\centering
\includegraphics[width=0.97\linewidth]{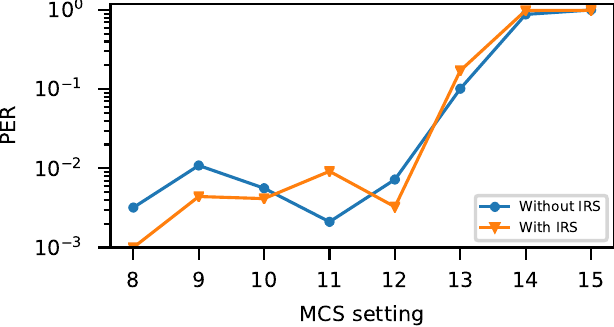}
\caption{Packet error rates over MCS setting without and with IRShield when a legitimate party receives packets from an anchor device. A higher MCS value indicates a higher data rate transmission.}
\label{fig:IRS_throughput}
\end{figure}

In Fig.~\ref{fig:IRS_throughput}, we plot the measured PER for each MCS~setting with and without IRShield being active. The first thing to note is that the PER increases with the MCS setting for both cases. As higher MCS~settings correspond to higher data rate transmissions with less immunity against noise, this indicates channel conditions with a limited signal-to-noise ratio. Apart from that, the key observation is that the channel obfuscation appears to cause only a negligible increase or decrease in PER. This is because our randomized IRS configuration approach will certainly both improve and degrade the wireless channel over time. Thus, as indicated by our experimental result, we believe that the average channel quality is not affected by channel obfuscation.%

\section{Discussion}\label{sec:discussion}

In the following, we discuss the experimental setup, our results and provide directions for future research.

\subsection{Experimental Setup and Results}
As our main goal was the evaluation of IRS-based channel obfuscation, we made some simplifications to the attack realization in order to facilitate experimentation. Notably, we thereby granted the adversary rather optimal conditions, which naturally is desirable for evaluation of a countermeasure.

First, we deployed a set of MIMO-capable \mbox{Wi-Fi} routers to act as the anchor and eavesdropper devices. Without loss of generality, the routers utilize standardized waveforms and can be considered as a substitute for any other \mbox{Wi-Fi} device. %
Further, the MIMO operation yields a total of $9$~spatial wireless channels in parallel from a single anchor device. %
Thus, the adversary takes advantage of spatial diversity~\cite{goldsmith_wireless_2005}, \ie, multiple different wireless channels, each containing information on the target area, to improve the adversarial motion detection~\cite{banerjeeViolating.2014}.

Another important aspect is the transmission rate of the anchor devices. As outlined in~\cite{zhu.2020}, typical \mbox{Wi-Fi} devices for use in living spaces transmit at $3$~to $257$~packets/sec during active operation. Likewise, in our setup, the anchors transmit at approx.~$70$~packets/sec. Thus, we constantly allow the adversary to observe the target area, even with substantial oversampling since human motion is rather slow when sampled at $70$~packets/sec. Further, the wireless traffic transmitted by our anchor devices does not carry meaningful application-driven data. However, this does not play a role for the attack which merely utilizes physical-layer observations, \ie, being independent of payload data and encryption.

We considered the threshold-based motion detection from~\cite{zhu.2020}. Here, long-term channel observations are used to find the detection threshold, assuming that the target area typically does not exhibit motion. For our experimental evaluation, other than in a realistic attack scenario, we granted the eavesdropper dedicated and well-behaved reference measurements without any movement in the target area.

Even under these clearly beneficial conditions for the adversary, IRShield almost completely defeats the scheme from the literature~\cite{zhu.2020}. Assuming the defense is known, we have also shown that the adversary in turn can adapt the threshold to restore some detection capabilities, albeit with strongly reduced success. This residual detectability is due to motion with very strong impact, \eg, within the LOS or close to the anchor, being stronger than the IRShield signal variation. %

\subsection{IRS Deployment}
The outlined probabilistic IRS configuration algorithm realizes channel obfuscation in a plug-and-play manner independently of anchor devices. Since the IRS directly interacts with wireless signals as they propagate, no integration efforts are required once the surface is deployed in proximity to an anchor device. While we used a host computer to implement Algorithm~\ref{alg:surface_configuration}, the IRS' on-board microcontroller can also directly implement the algorithm, allowing fully stand-alone operation. Note that, however, the randomized IRS configurations are sub-optimal as they do not maximize the IRS impact, especially for higher distances between the IRS and the anchor devices. To this end, we emphasize that IRSs are being considered for future 6G wireless communication systems with potential for integration into walls and furniture~\cite{wuIntelligentReflectingSurface2021}. Thus, the assumption of IRS deployments in proximity to anchor devices indeed is well justified.

The effectiveness of channel obfuscation is rooted in the used IRS hardware. The IRS is commonly considered as low-complexity device that can be built at low cost, using standard PCB technology. The IRS deployed in our experiments is a prototypical device which leaves room for hardware improvements. For instance, currently only \SI{5}{\GHz} frequency bands are covered and signals are reflected with a certain loss. Thus, our results may be further improved as the technology evolves. %

\subsection{Bypassing IRShield}
Advanced adversaries may attempt to remove or suppress the effect of IRShield. %
By using antenna arrays for beamforming (or directional antennas), signal components can be distinguished in the angular domain. However, for IRShield, this is not possible due to the proximity of the IRS to the anchor, i.e., the IRS signals emanate from the same direction as the direct signals. Further, as indicated by our experimental results, it appears that the IRS yields channel variation in all directions. Another option would be time-domain analysis where individual multipath signal components are distinguishable by their respective time of arrival at the receiver. While RF transmitters often do not provide sufficient bandwidth for this approach to work properly (time resolution scales inversely with bandwidth), the IRShield signals further encounter similar delay spreads compared to the rest of the propagation environment.

Without changing signal domains, adversaries could also strive to leverage the specific behavior of the IRS configuration algorithm. For instance, it would be conceivable to combine consecutive channel measurements before and after the inversion step of Algorithm~\ref{alg:surface_configuration}. Due to the sign change, the IRS contribution would vanish. In practice, the bottleneck to this approach are stable phase-coherent channel measurements over time which are difficult to obtain due to the asynchronous distant transmitter and receiver. In fact, signal phase of commodity radios faces various severe random distortions~\cite{ma.2019}. Still, an alternative algorithm design could replace the inversion by a second random step, \eg, where only a certain fraction of randomly selected elements is flipped.

\subsection{Further Applications of IRShield}

As \mbox{Wi-Fi} devices are particularly widespread in private and public spaces, we used IEEE~802.11n \mbox{Wi-Fi} to confirm the potential of IRS for channel obfuscation. However, in some cases, an adversary may leverage other wireless standards for sensing. Since the IRS operates directly on the physical layer and affects the wireless channel, it is independent of particular waveforms or standards. Thus, IRShield generally is not tied to \mbox{Wi-Fi}. The experimental verification of IRShield for other wireless standards is left for future work.

We examined channel obfuscation as a countermeasure particularly geared towards adversarial (human) motion detection. However, beyond that, applications such as gesture recognition, identification of individuals, imaging, vital sign monitoring, or keystroke recognition~\cite{ma.2019} have been reported and can have far-reaching privacy implications. We believe that the intricate signal processing methods employed for these applications could be susceptible against signal variations induced by the IRS. Therefore, IRShield may also protect against other classes of wireless sensing, which can be confirmed through further experimentation.

Finally, we would like to point out that the IRS could also be used to corrupt legitimate wireless sensing applications, \eg, for motion-based intrusion detection~\cite{linksysAware}. However, such scenarios are beyond the scope of this work.%

\subsection{Limitations}
IRShield significantly hampers adversarial motion detection but cannot fully prevent it in some cases. As our results show (see Fig.~\ref{fig:heatmaps_LOS_maxref} in Appendix~\ref{sec:heatmaps_reduced_threshold}), this effect occurs when motion takes place close to anchors and within the LOS. That is, environmental variation can cause RF signal fluctuation which is stronger than IRS-induced signal variation, yielding a residual detectability. This limitation of IRShield is rooted in the IRS' finite impact on the signal propagation: The IRS is purely passive, \ie, it reflects signals but cannot apply amplification. Further, the IRS surface configuration requires optimization to achieve maximum signal power -- a requirement not met by the randomized surface configurations of IRShield. To enhance the impact, the defender could increase the IRS size or reduce the distance to anchors. Further, at the cost of losing device independence, IRShield could also cooperate with legitimate devices to optimize IRS configurations. Another simpler option for the defender is strategic anchor placement. As we have shown, anchor devices close to the eavesdropper are less sensitive against motion. By taking advantage of limited public accessibility of the target area, the defender can effectively prevent motion to take place within the eavesdropper's LOS to further diminish the adversary's success. Finally, the discussed residual detectability is specific to motion sensing. For other wireless sensing tasks based on weaker signal features, the relative impact of IRShield would rise.

Another limitation of IRShield is its undirected effect that would interfere with legitimate wireless sensing applications. To resolve this, IRShield would need to interact with the legitimate parties, \ie, to estimate the IRS channels~\cite{Zheng.2020} and communicate IRS configurations. Based on this information, a legitimate receiver could attempt to remove the effect of IRShield.

\subsection{Future Work}
Our findings may serve as a starting point for future work that could investigate, for instance, the following aspects.

\Paragraph{Optimization of IRS configuration}
We proposed a probabilistic IRS configuration algorithm for IRShield. While we already obtained satisfactory results, we believe that there is still room for improvements, \eg, by randomizing state transitions of the algorithm or by switching between predefined patterns. Currently, IRShield operates independently of the anchor device. The next logical step would be to incorporate information from the anchor devices and accordingly, adjust the IRS, \eg, by analyzing channel estimates.

\Paragraph{Beamforming}
More recent \mbox{Wi-Fi} devices, \eg, for IEEE~802.11ac, utilize beamforming. Thus, wireless signals are no longer transmitted to all directions (including the eavesdropper) but rather directed towards the intended receiver. Hence, an eavesdropper will receive weaker signals~\cite{antonioliPracticalEvaluationPassive2018} which likely reduces the adversarial sensing capabilities. More work is required to investigate whether beamforming in conjunction with IRShield further degrades the adversary's capabilities.

\Paragraph{Further testing and adversary strategies}
We showed that \mbox{IRShield} defeats existing adversarial motion-detection strategies in an exemplary test scenario. However, a powerful adversary may use multiple receivers or apply different signal processing and statistical classification methods. Future work should further investigate the performance in more elaborate settings, \eg, with more anchor devices, varying room sizes, multiple rooms, and motion of multiple users.

\Paragraph{Legitimate sensing}
The coexistence of IRShield and legitimate wireless sensing applications could also be further explored. In particular, IRShield is likely to affect legitimate sensing capabilities. Therefore, it would be interesting to explore solutions to leave authorized parties unaffected or allow them to remove the effect of IRShield.

\section{Related Work}\label{ref:relatedwork}

In this section, we summarize relevant literature on privacy violations caused by sniffing wireless devices and report related work on wireless sensing. We also outline how our work differs from previous proposals for countermeasures. 

With a particular focus on privacy violations, Banerjee~\etal~\cite{banerjeeViolating.2014} and Zhu~\etal~\cite{zhu.2020} studied \mbox{Wi-Fi}-based physical-layer attacks to infer human motion from channel observations by eavesdropping packets. In both papers, the authors utilized a sliding-window standard deviation or variance as a measure of temporal channel variation. While \cite{banerjeeViolating.2014} focused on detecting LOS crossings, \cite{zhu.2020} pursues monitoring of entire environments for motion, including differentiation of motion in specific parts of the target environment. The authors of~\cite{banerjeeViolating.2014} employed \mbox{Wi-Fi} routers with MIMO functionality to implement the attack and~\cite{zhu.2020} presented a smartphone-based implementation.

To counteract unauthorized wireless sensing, different types of countermeasures have been considered. Qiao~\etal proposed PhyCloak~\cite{qiao.2016}, a full-duplex relay node to introduce channel variation. Wijewardena~\etal~\cite{wijewardena.2020} put forward a game-theoretic framework to trade the privacy loss against the quality of service of the wireless communication. %
Jiao~\etal~\cite{jiao.2021} presented an extension of the \mbox{Wi-Fi} transmitter module of openwifi which is a full-stack software-defined radio \mbox{Wi-Fi} implementation~\cite{openwifigithub}. Using a reconfigurable digital filter, an artificial channel response is imposed on transmitted signals to disguise the actual environment-dependent channel response.

Apart from motion sensing, there is a large body of work on \mbox{Wi-Fi} sensing applications as surveyed by Ma~\etal~\cite{ma.2019}. For instance, \mbox{Wi-Fi} signals have been employed for gesture recognition~\cite{AbdelnasserWiGest2015}, identification of individuals~\cite{Li.2016}, imaging~\cite{HuangWiFiImaging2014}, vital sign monitoring~\cite{LiuVitalSign2015}, or keystroke recognition~\cite{AliKeystroke2015}. Shifting the scope of wireless sensing, Camurati~\etal~\cite{camurati.2018} have shown that RF signals from Bluetooth and \mbox{Wi-Fi} chipsets can be utilized for side-channel attacks against on-chip cryptographic hardware implementations from rather large distances. %

\Paragraph{Differentiation from previous work}
A conceptually closely related work is PhyCloak introduced by Qiao~\etal~\cite{qiao.2016}. Using a full-duplex radio, legitimate RF signals are received and re-transmitted while applying randomized signal changes to hamper adversarial sensing. However, full-duplex requires specialized and highly complex radio hardware to achieve self-interference cancellation in order to allow for simultaneously receiving and transmitting on the same frequency. Another proposal by~Yao~\etal~\cite{yao.2018} also receives and re-transmits legitimate signals but uses full-duplex by means of a directional transmit antenna. To introduce randomized changes to the adversarial observations, however, the setup employs several moving parts, \eg, a motor to rotate the antenna and a fast fan. While IRShield follows the general receive-and-re-transmit rationale, our approach works entirely different. Crucially, we utilize an IRS to unify reception, signal alteration, and re-transmission of signals into digitally adjustable reflection. In contrast to \cite{qiao.2016} and \cite{yao.2018}, the IRS is inherently capable of full-duplex operation and does not require moving parts.

Along with their attack implementation, Zhu~\etal~\cite{zhu.2020} proposed a countermeasure based on a fake access point to inject cover packets to mimic legitimate traffic. As a result, the eavesdropper observes a mix of packets transmitted by the victim devices and the fake access point, thus reducing the adversarial sensing capability. We believe the adversary can easily defeat this strategy since the channel from the fake access point is clearly distinguishable due to the location-dependence of wireless channels~\cite{goldsmith_wireless_2005}. Another interesting approach put forward by Wijewardena~\etal~\cite{wijewardena.2020} is based on strategically randomizing transmit power or turning off devices completely. However, both strategies are based on changing the allocation of the channel and therefore, face a trade-off between the quality of service of wireless communication and reduction of adversarial sensing capabilities. We emphasize that IRShield rather changes the wireless channel itself instead of its allocation. Moreover, IRShield is independent of the existing equipment, \ie, it neither requires adjustments of transmit power or timing, nor does it need to be matched to the legitimate signals.

\section{Conclusion}\label{sec:conclusion}
In this paper, we introduced and experimentally examined IRShield as a plug-and-play countermeasure against adversarial motion detection from passive eavesdropping of wireless signals. By deploying an IRS, we partly randomize the wireless propagation environment to introduce random variations which obfuscate the adversary's signal observations. To this end, we presented an extensive experimental investigation to characterize channel obfuscation. Beginning with human motion detection, we have shown that our scheme successfully diminishes the adversary's success. We then systematically studied spatial aspects of the attack and the countermeasure and discussed how the adversary could adapt to the defense. Notably, we found that our scheme lowered detection rates to~\SI{5}{\percent} or less for a state-of-the-art attack. In certain cases, it renders motion detection largely infeasible, regardless of the adversary strategy. Furthermore, we investigated IRS parameters such as size, distance, and orientation and showed that IRShield does not affect the wireless communication performance.

\section*{Acknowledgements}
We thank Dr. Christian Zenger and PHYSEC GmbH for their support with the IRS prototypes. We thank Simon Tewes for his help with setting up the \mbox{Wi-Fi} routers. This work was in part funded by the German Federal Ministry of Education and Research~(BMBF) within the project MetaSEC~(Grant 16KIS1234K) and by the Deutsche Forschungsgemeinschaft~(DFG, German Research Foundation) under Germany’s Excellence Strategy - EXC 2092 CaSa - 390781972.

\bibliographystyle{IEEEtranS}
\bibliography{sensing_defense_refs}

\appendices

\section{Algorithm Parameters}
\label{sec:algorithm_parameters}
In Section~\ref{sec:irshield}, we introduced Algorithm~\ref{alg:surface_configuration} to generate the IRS configurations for IRShield. Here we provide additional details on the algorithm parametrization.

\begin{figure}[h!]
\centering
\includegraphics[width=1\linewidth]{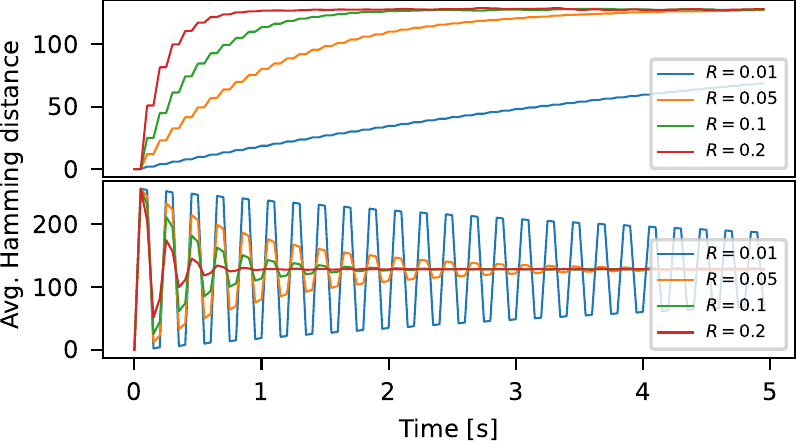}%
\caption{Average Hamming distances of IRS configurations w.r.t. $t=0$, without (top) and with (bottom) inversion step.}
\label{fig:hamming_vs_time}
\end{figure}

\Paragraph{Progression rate $R$}
The progression rate $R$ affects the decorrelation speed of IRS configurations. In particular, $R$ determines the number of randomly selected IRS elements $\ceil*{R \cdot M}$ which are flipped within the algorithm's randomization step. Therefore, $R \in [0,0.5]$. For various values of~$R$, we simulate ensembles of IRS configurations over time with $P_{\mathrm{hold}} = 0$, $M=256$ IRS elements, and an update rate of $20$~configurations/sec. In Fig.~\ref{fig:hamming_vs_time}, we plot the average Hamming distance over time to the initial configuration at $t=0$. For the sake of simplicity, the upper plot shows the behavior of Algorithm~\ref{alg:surface_configuration} with the inversion step being disabled (having no effect). The lower plot is obtained using the unmodified algorithm. %

\begin{figure}[h!]
\centering
\includegraphics[width=1\linewidth]{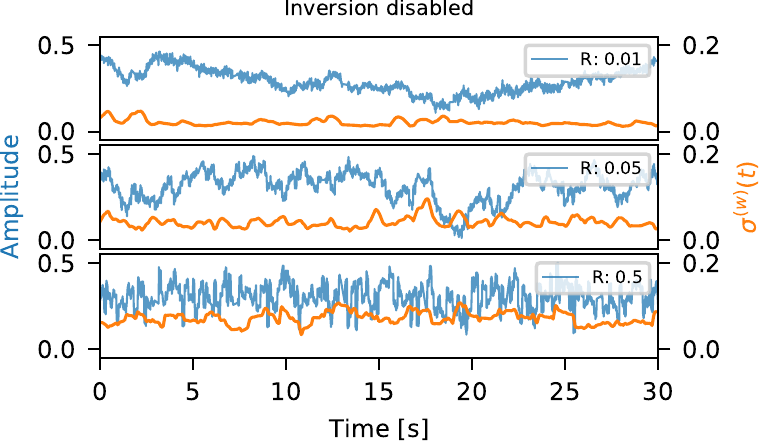}%
\caption{Time series of a single OFDM subcarrier from \mbox{Wi-Fi} measurements and \SI{1}{\s}~sliding-window standard deviation for various values of $R$. The inversion step of Algorithm~\ref{alg:surface_configuration} is bypassed.}
\label{fig:multi_R_csi_timeseries_noflip}
\end{figure}

\begin{figure}[h!]
\centering
\includegraphics[width=1\linewidth]{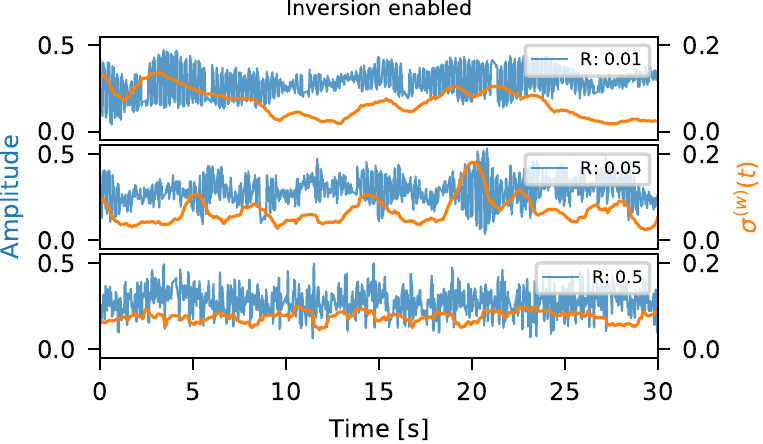}%
\caption{Time series of a single OFDM subcarrier from \mbox{Wi-Fi} measurements and \SI{1}{\s}~sliding-window standard deviation for various values of $R$.}
\label{fig:multi_R_csi_timeseries_withflip}
\end{figure}

Since the channel via the IRS, $\bm{H}_{\textrm{IRS}}(k,t)$, is a function of the IRS configuration, $R$ directly affects the decorrelation speed of the time series of IRS channels. Intuitively, a change of few IRS elements, \eg, \SI{5}{\percent}, yields a small change in the channel. On the other hand, changing many IRS elements, \eg, \SI{50}{\percent}, is likely to result in a larger change. Thus, we expect that reducing $R$ increases the smoothness of the time series of IRS channels. To experimentally assess this effect, we collect \mbox{Wi-Fi} CSI measurements using the setup described in Section~\ref{sec:router_setup}. We place the IRS at a distance of~\SI{30}{\cm} to the anchor and vary $R$. Similar to the simulation, we use $P_{\mathrm{hold}} = 0$, $M=256$ IRS elements, and an update rate of $20$~configurations/sec. We plot the amplitude of a single OFDM subcarrier over time for exemplary values of $R$ with (Fig.~\ref{fig:multi_R_csi_timeseries_noflip}) and without (Fig.~\ref{fig:multi_R_csi_timeseries_withflip}) the inversion step of Algorithm~\ref{alg:surface_configuration} being bypassed. From both figures, we can see the algorithm's desired to effect to vary the IRS signal amplitude over time. As expected, it is evident that $R$ controls the speed of these variations. Apart from $R$, the effect of the inversion step can be observed in Fig.~\ref{fig:multi_R_csi_timeseries_withflip}, laying the foundation for an enhanced sliding standard deviation.

$R$ should be chosen low enough such that strong IRS amplitudes persist during the adversary's sliding standard deviation window to create excursions within $\bar{\sigma}^{(w)}(t)$. Conversely, when the IRS changes too fast, single strong IRS amplitude deviations will have less impact on $\bar{\sigma}^{(w)}(t)$. This can be observed in Fig.~\ref{fig:multi_R_csi_timeseries_withflip}, where we also plot the sliding standard deviations of the CSI time series.

\begin{figure}[h!]
\centering
\includegraphics[width=1\linewidth]{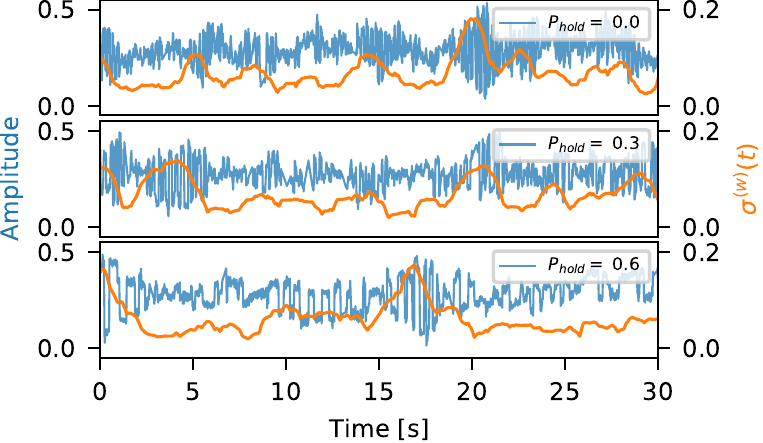}%
\caption{Time series of a single OFDM subcarrier from \mbox{Wi-Fi} measurements and \SI{1}{\s}~sliding-window standard deviation for various values of $P_{\mathrm{hold}}$.}
\label{fig:phold_figure}
\end{figure}

\Paragraph{Hold probability $P_{\mathrm{hold}}$}
In the previous paragraph, we considered Algorithm~\ref{alg:surface_configuration} with $P_{\mathrm{hold}}=0$. In this case, the algorithm is strictly scheduled according to the IRS refresh rate which we set to $20$~configurations/sec. With $P_{\mathrm{hold}} > 0$, the algorithm is halted in a probabilistic manner, randomizing the timing of the state transitions and scaling the average IRS refresh rate by the factor $P_{\mathrm{hold}}$. In turn, short pauses are inserted into the IRS operation, such that $\bar{\sigma}^{(w)}(t)$ covers a wider range. To investigate the effect, we use the experimental setup outlined in the previous paragraph and collect \mbox{Wi-Fi} CSI measurements for $R=0.05$ and varying values of $P_{\mathrm{hold}}$. We plot the amplitude of a single OFDM subcarrier and the sliding standard deviation over time in Fig.~\ref{fig:phold_figure}. From the bottom plot, showing $P_{hold} = 0.6$, we can clearly see that the timing of the IRS-induced channel variation becomes increasingly non-uniform.

\Paragraph{Choice of parameters}
Having outlined the effect of $R$ and $P_{\mathrm{hold}}$, we next seek to determine values for $R$ and $P_{\mathrm{hold}}$. Based on the considerations from Section~\ref{sec:intro_irs_obfuscation}, we use the following metrics to assess the algorithm parametrization:
\begin{itemize}
\item IRShield should disturb the adversary's reference measurement to yield an overly high motion detection threshold, cf. Eq.~(\ref{eq:threshold_motion}). Thus, the median and MAD of $\bar{\sigma}^{(w)}(t)$ should be maximized.
\item IRShield should produce strong time-varying fluctuations within the adversarial observation. Thus, the signal energy (the Euclidean norm) of $\bar{\sigma}^{(w)}(t)$ should be maximized.
\item The adversarial observation with IRShield should exhibit temporal characteristics similar to actual human motion, \cf~Fig.~\ref{fig:motion_sensing_attack1}, where strong signal variation occurs for durations of \SI{1}~--~\SI{2}{\s}. We assess the decorrelation speed by means of the coherence time of $\bar{\sigma}^{(w)}(t)$ as the time its normalized autocorrelation lies above $0.5$.
\end{itemize}

\begin{figure}[h!]
\centering
\includegraphics[width=1\linewidth]{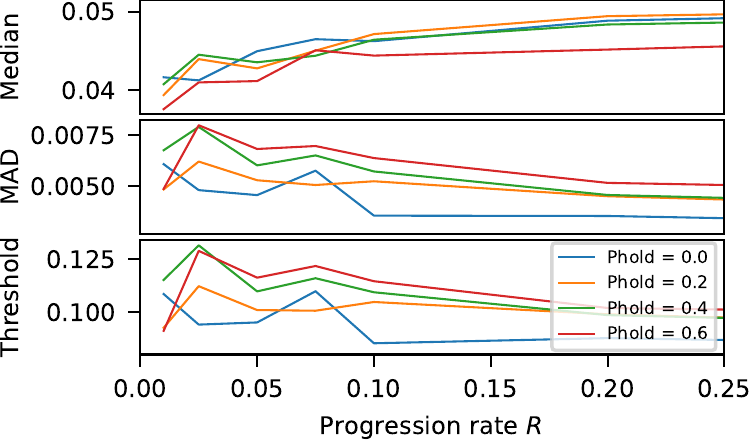}%
\caption{Median, MAD of $\bar{\sigma}^{(w)}(t)$ and the resulting adversarial motion detection threshold.}
\label{fig:threshold_vs_R_P}
\end{figure}

\begin{figure}[h!]
\centering
\includegraphics[width=1\linewidth]{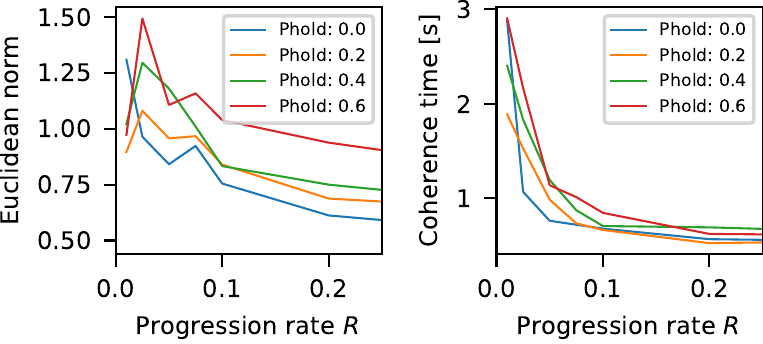}%
\caption{Euclidean norm (left) and coherence time (right) of $\bar{\sigma}^{(w)}(t)$.}
\label{fig:l2_vs_R_P}
\end{figure}

As we can see from Figs.~\ref{fig:threshold_vs_R_P} and~\ref{fig:l2_vs_R_P}, $P_{\mathrm{hold}}$ should be at least $0.4$, while $R$ should be between~$0.025$ and~$0.05$. Thereby, the adversarial motion detection threshold and the Euclidean norm of $\bar{\sigma}^{(w)}(t)$ are maximized while the coherence time lies within the desired range of approx.~\SI{1}{}~to~\SI{2}{\s}.

\section{Systematic Coverage Experimental Setup}
\label{sec:coverage_setup}
\begin{figure}[h!]
\centering
\includegraphics[width=1\linewidth]{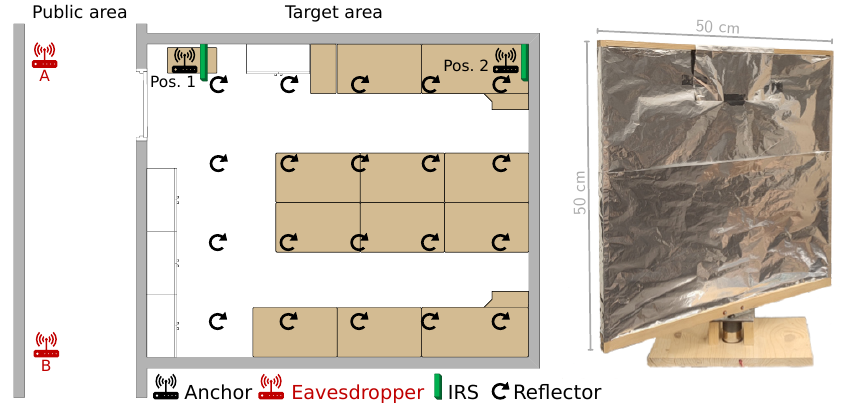}%
\caption{Measurement setup of the coverage test, indicating the $20$~rotating reflector positions~(left) and the rotating reflector on a wooden frame~(right).}
\label{fig:systematic_experiment_setup}
\end{figure}

The experimental setup for the systematic coverage tests as described in Section~\ref{sec:heatmap} is depicted in Fig.~\ref{fig:systematic_experiment_setup}.

\section{Heatmaps with Reduced Threshold}
\label{sec:heatmaps_reduced_threshold}

\begin{figure}[!htb]
\centering
\subfloat[Anchor $1$, Eve $A$, without IRS]{\includegraphics[width=0.49\linewidth]{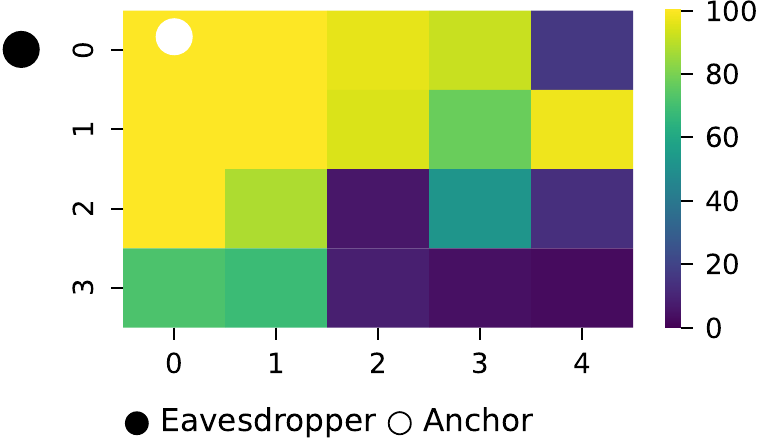}%
} \hfill
\subfloat[Anchor $1$, Eve $A$, with IRS]{\includegraphics[width=0.49\linewidth]{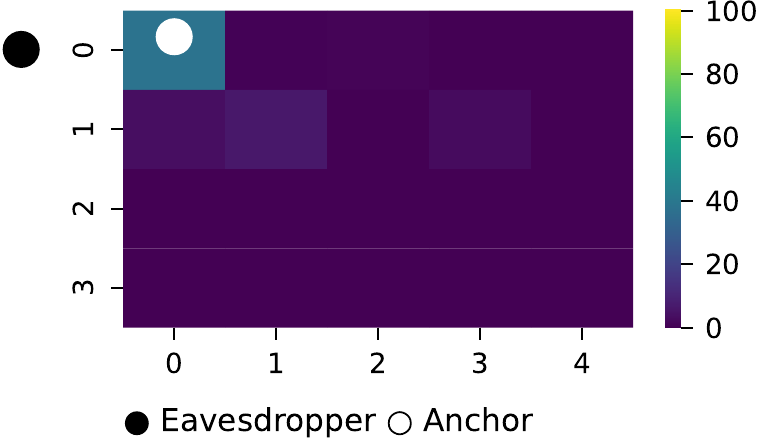}%
}\\
\subfloat[Anchor $1$, Eve $B$, without IRS]{\includegraphics[width=0.49\linewidth]{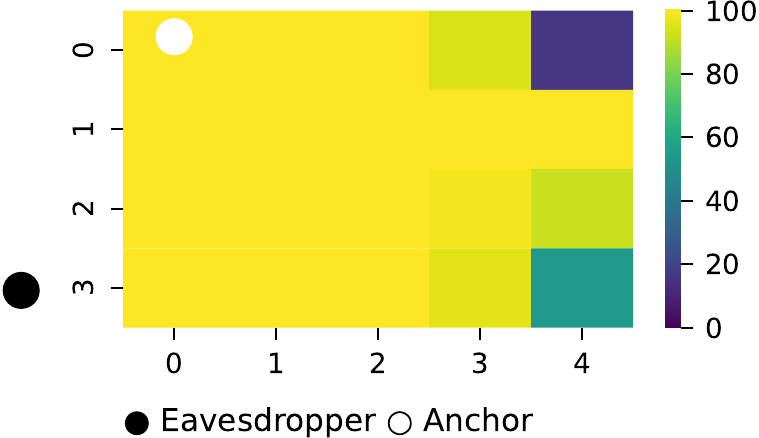}%
} \hfill
\subfloat[Anchor $1$, Eve $B$, with IRS]{\includegraphics[width=0.49\linewidth]{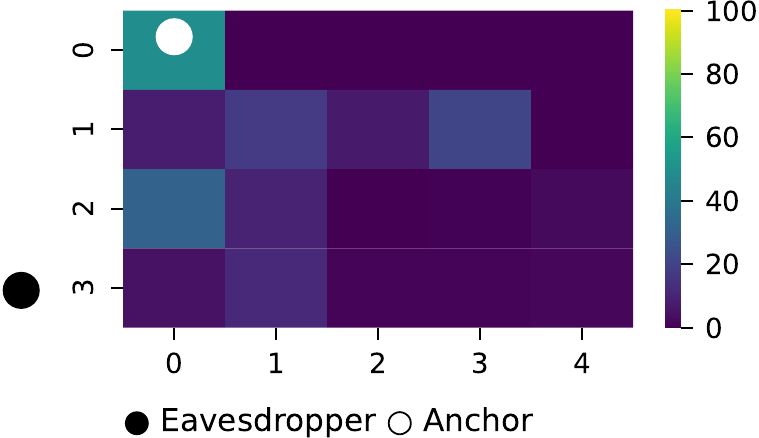}%
}
\caption{Spatial distribution of detection rates for anchor position $1$ with lowest motion-detection threshold yielding a false-positive rate of $0$.} 
\label{fig:heatmaps_NLOS_maxref}
\end{figure}

\begin{figure}[!htb]
\centering
\subfloat[Anchor $2$, Eve $A$, without IRS]{\includegraphics[width=0.49\linewidth]{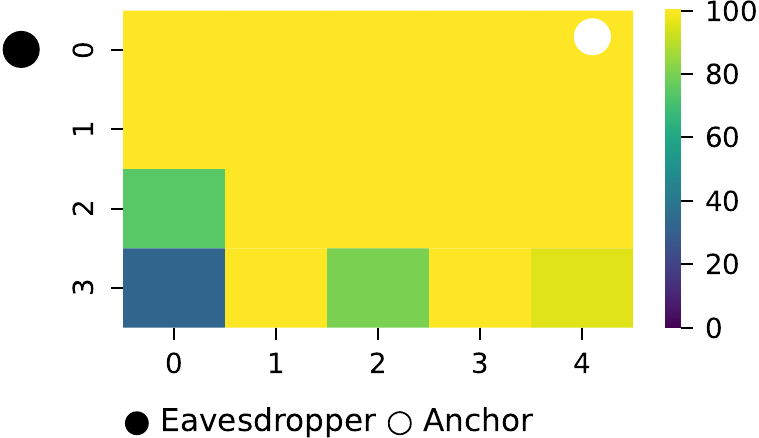}%
} \hfill
\subfloat[Anchor $2$, Eve $A$, with IRS]{\includegraphics[width=0.49\linewidth]{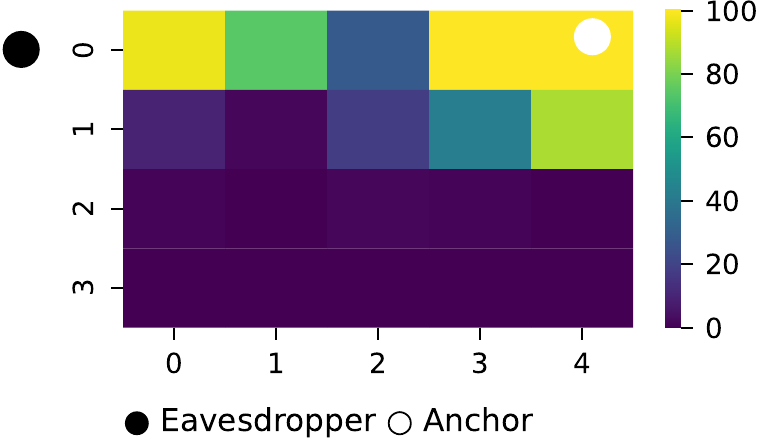}%
}\\
\subfloat[Anchor $2$, Eve $B$, without IRS]{\includegraphics[width=0.49\linewidth]{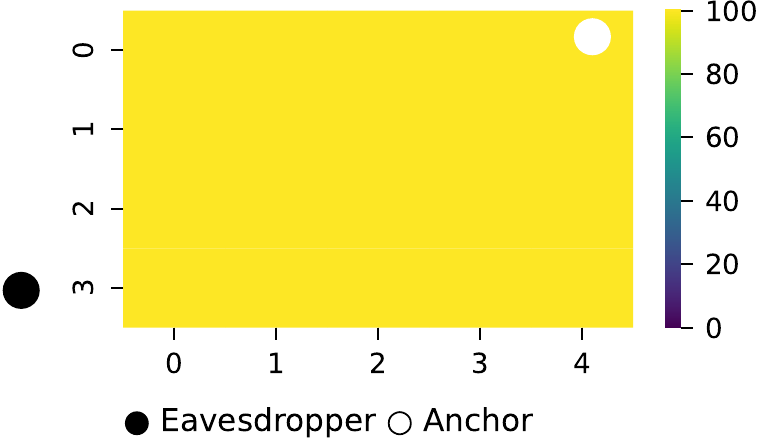}%
} \hfill
\subfloat[Anchor $2$, Eve $B$, with IRS]{\includegraphics[width=0.49\linewidth]{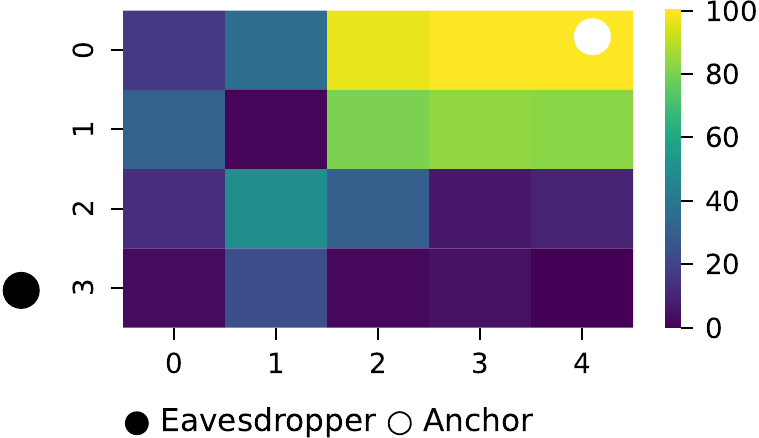}%
}
\caption{Spatial distribution of detection rates for anchor position $2$ with lowest motion-detection threshold yielding a false-positive rate of $0$.} 
\label{fig:heatmaps_LOS_maxref}
\end{figure}

In Section~\ref{sec:heatmap}, we systematically studied the spatial performance of adversarial motion sensing with and without IRShield. Complementing Fig.~\ref{fig:heatmaps_NLOS_alexa} and Fig.~\ref{fig:heatmaps_LOS_alexa}, Fig.~\ref{fig:heatmaps_NLOS_maxref} and Fig.~\ref{fig:heatmaps_LOS_maxref} show the spatial distribution of motion detection rates with the maximum of the reference measurement being used as motion-detection threshold. Thus, the detection rates were obtained with the lowest threshold resulting in a FPR of $0$ (at least for the observed reference duration). This corresponds to the first row of Table~\ref{tab:detection_rates_text}, indicated by $\mathrm{max}_t\left\{\cdot\right\}$.

\end{document}